\documentclass[twocolumn,showpacs,
preprintnumbers,amsmath,amssymb,floatfix]{revtex4-1}

\usepackage{natbib}
\usepackage{float}

\DeclareMathAlphabet{\mathpzc}{OT1}{pzc}{m}{it}

\usepackage{graphicx}
\usepackage{amsmath,amsfonts,amssymb}
\usepackage{graphicx}
\usepackage{tabularx}
\usepackage{color}
\usepackage{bbold}
\usepackage{braket}

\usepackage[normalem]{ulem}
\usepackage{todonotes}
\usepackage[bookmarks,
            breaklinks,
            pdfstartview=FitH, 
            pdftitle={},
            pdfauthor={},
            pdfkeywords={}
            ]{hyperref}
\usepackage{ulem}
\hypersetup{
  colorlinks   = true,  
  urlcolor     = blue,  
  linkcolor    = blue,  
  citecolor    = red    
}

\usepackage[margin=1.in]{geometry}
\usepackage{graphicx}
\usepackage{amsmath}
\usepackage{amssymb}
\usepackage{mathtools}
\usepackage{amsthm}
\usepackage{tikz-cd}
\usepackage{subfigure}
\usepackage{booktabs}

\newcommand{\bb}[1]{\mathbf{#1}}

\DeclareUnicodeCharacter{2212}{-}
\DeclareUnicodeCharacter{02B9}{-}

\begin{document}

\title{Multichannel effects in the Efimov regime from broad to narrow Feshbach resonances}
\author{T. Secker}
\affiliation{Eindhoven University of Technology, P.~O.~Box 513, 5600 MB Eindhoven, The Netherlands}
\author{D. J. M. Ahmed-Braun}
\affiliation{Eindhoven University of Technology, P.~O.~Box 513, 5600 MB Eindhoven, The Netherlands}
\author{P. M. A. Mestrom}
\affiliation{Eindhoven University of Technology, P.~O.~Box 513, 5600 MB Eindhoven, The Netherlands}
\author{S. J. J. M. F. Kokkelmans}
\affiliation{Eindhoven University of Technology, P.~O.~Box 513, 5600 MB Eindhoven, The Netherlands}

\date{\today}


\begin{abstract}

We study Efimov physics of three identical bosons with pairwise multichannel interactions for Feshbach resonances of adjustable width.
We find that the two-body multichannel nature of the interaction can affect the universal three-body spectrum, especially for resonances of intermediate width.
The shifts in this universal spectrum are caused by trimer states in the closed interaction channels that couple to the universal Efimov states.
However, in the narrow resonance limit we find that the Efimov spectrum is set by the resonance width parameter $r^*$ only independent of the interaction potential considered.
In the broad resonance limit all excited Efimov trimer energies approach the ones from the corresponding single-channel system for the scenarios investigated.

\end{abstract}

\maketitle

\section{Introduction}
As originally analyzed by Efimov, three particles that interact with resonant pairwise interactions show
universal behavior independent of the details of the interaction potentials \cite{Efimov:1970}.
Whereas Efimov first analyzed this effect in the context of nuclear physics, over the last decades ultracold alkali atoms have proven to be an ideal platform to study Efimov physics experimentally for Bose gases \cite{Knoop:2009,Zaccanti:2009,Gross:2009,Pollack:2009,Gross:2010}, Fermi gases \cite{Ottenstein:2008,Lompe:2010,Huckans:2009,Williams:2009,Nakajima:2011} and mixtures \cite{Barontini:2009,Pires:2014,Pires:2014_2,Tung:2014,Ulmanis:2015,Johansen:2017,Ulmanis:2016,Ulmanis:2016_2,Wacker:2016}. 
In these atomic systems, the pairwise interaction can be tuned into the resonant regime close to Feshbach resonances by applying external magnetic fields \cite{Chin:2010}. 
This tunability is a consequence of Zeeman shifts in the atomic hyperfine states of the individual atoms.
Combinations of these internal hyperfine states on the two and three atom level form the different scattering channels of the system which are coupled by a multichannel interaction potential when the atoms approach each other.

The strength of the interaction in the underlying two-body system can be parametrized by the $s$-wave scattering length $a$ in the ultracold regime. 
The behavior of $a$ near a Feshbach resonance  can then be characterized by a background scattering length $a_\mathrm{bg}$ and a resonance width parameter $r^*$ \cite{Chin:2010}. 
The parameter $r^*$ is related to the width of the resonance in magnetic field $\Delta B = \hbar^2 / m \, d\mu \, a_\mathrm{bg} \, r^*$, with $m$ the mass of the atoms and $d \mu$ the magnetic moment of the bound state associated with the resonance.
Large values of $r^*$ therefore correspond to narrow Feshbach resonances, whereas small values correspond to broad ones.
The term $\Delta B$ can be determined from the magnetic field dependence of $a$
\begin{equation}
a(B) = a_\mathrm{bg}\left(1-\frac{\Delta B}{B - B_0} \right) \, ,
\end{equation}
where $B$ and $B_0$ denote the magnetic field strength and the resonance position in magnetic field, respectively.

The universal Efimov regime is characterized by large absolute values of the scattering length, which diverges on resonance ($|a| \gg r_\mathrm{vdW}$), where $r_\mathrm{vdW}=(m C_6 / \hbar^2)^{1/4} / 2$ is the range associated with the $-C_6/r^6$ van der Waals tail of the atomic interaction.
In the universal Efimov regime a single three-body parameter determines the location of three-body features, e.g. the binding energies $E_n$ of an infinite sequence of weakly bound three-body states, referred to as Efimov trimers, that emerge successively when the interaction is tuned to resonance.
The binding energies follow the universal scaling relation, 
$E_{n+1}/E_{n} = e^{-2\pi/s_{0}}$ with $s_{0}\approx 1.00624$ for identical bosons \cite{Efimov:1970,Braaten:2006}.
The three-body parameter is often determined by the scattering length value $a_-^{(0)}$ at which the lowest Efimov trimer state meets the three-body continuum and leads to an Efimov resonance.

The three-body parameter $a_-^{(0)}$ has been measured for many species and Feshbach resonances \cite{Kraemer:2006,Pollack:2009,Gross:2009,Gross:2010,Zaccanti:2009,Wild:2012,Ferlaino:2011,Berninger:2012}.
Surprisingly, most early experiments for identical bosons found the three-body parameter to be close to $a_-^{(0)} / r_{\mathrm{vdW}} \approx -9 $ over different species and Feshbach resonances \cite{Kraemer:2006,Gross:2009,Gross:2010,Wild:2012,Ferlaino:2011,Berninger:2012}.
Following the experiments, this universal value of $a_-^{(0)}$ could be theoretically explained \cite{Wang:2012,Naidon:2014a,Naidon:2014l} relying on single-channel models for the interaction potential.
However, such single-channel models cannot correctly describe the two-body physics close to narrow Feshbach resonances.

For narrow resonances $a_-^{(0)}$ can deviate from the universal value as has been predicted theoretically \cite{Petrov:2004,Gogolin:2008,Schmidt:2012,Langmack:2018,Sorensen:2012,Chapurin:2019,Secker:2020b} and observed experimentally \cite{Roy:2013,Chapurin:2019, Gross:2011, Dyke:2013}.
However, agreement between theory and experiment has been achieved only in a few instances \cite{Chapurin:2019,Secker:2020b}.
The deviation in $a_-^{(0)}$ from the universal value thus indicates the importance of multichannel effects for narrow resonances.
To represent narrow resonances on the two-body level simple two-channel models can be used to correctly account for the resonance width parameter $r^*$.
Generalizing those models to the three-body case resulted in studies of $a_-^{(0)}$ as a function of $r^*$ \cite{Petrov:2004,Gogolin:2008,Schmidt:2012,Sorensen:2012}, as well as of both $r^*$ and $a_\mathrm{bg}$ \cite{Langmack:2018}.
All those models find a dependence of $a_-^{(0)}$ on $r^*$, however opposite trends for the behavior of $|a_-^{(0)}|$ have been obtained.
On the one hand $|a_-^{(0)}|$ can increase with increasing $r^*$ reaching a narrow resonance limit ($r^* \rightarrow \infty$) where $r^*$ sets the three-body universal regime as well as the three-body parameter \cite{Petrov:2004,Gogolin:2008,Schmidt:2012} or on the other hand $|a_-^{(0)}|$ can decrease with increasing $r^*$ \cite{Sorensen:2012}.
Interestingly both increasing \cite{Roy:2013,Chapurin:2019} and decreasing \cite{Gross:2011,Dyke:2013} trends have also been observed experimentally.
One of the major differences in the models that predict opposite behavior is the way the two-body multichannel structure is embedded into 
the three-body multichannel space.
This indicates the importance of an accurate representation of the multichannel structure of the three-atom system.
 
On the two-body level, a symmetric spin channel is in general either of the form $| \mathpzc{c} \mathpzc{c}  \rangle $ or $ | \mathpzc{c} \mathpzc{c}' \rangle_S = (| \mathpzc{c} \mathpzc{c}' \rangle + | \mathpzc{c}' \mathpzc{c} \rangle) / \sqrt{2} $, with $\mathpzc{c}$ and $\mathpzc{c}'$ labeling the different internal spin states of the individual atoms.
For identical bosons such symmetric spin channels need to be considered when combined with even partial wave components of the interaction.
In this paper we analyze both realizations $| \mathpzc{c} \mathpzc{c}  \rangle $ and $ | \mathpzc{c} \mathpzc{c}' \rangle_S $ for the closed channel and are thereby extending earlier studies \cite{Langmack:2018,Schmidt:2012,Gogolin:2008,Petrov:2004} to the $ | \mathpzc{c} \mathpzc{c}' \rangle_S $ scenario. 
We consider pairwise separable interaction potentials that have been used to study the dynamics of the many-body system \cite{Colussi:2018, Musolino:2019} and generalize them to a multi-channel interaction. 
We find that the realization of the two-channel model in spin space affects the Efimov spectrum in the broad and especially intermediate resonance width regime, leading to both increasing and decreasing trends of $|a_-|$ depending on the realization, while the narrow resonance limit can be recovered for $r^* \rightarrow \infty$ in all considered cases. 
Additionally, we find the narrow resonance limit to be independent of the form of the interaction we consider.

The paper is outlined as follows. 
In Sec. \ref{sec:Three-body bound states in a multichannel setting}, we start with the analysis of the three-body bound state equations in momentum representation for a general multichannel system. 
To keep the system clean whilst retaining the multichannel characteristics, we proceed to analyze a simple two-channel model in Sec. \ref{sec:Two-channel model system}, where we restrict ourselves to the case of separable $s$-wave interactions and two internal spin states $| a \rangle$ and $| b \rangle$ per particle. Here, we distinguish between the two different closed-channel realizations $\ket{bb}$ and $\ket{ab}_S$ we mentioned earlier. 
The two-body details are discussed in section \ref{sec:Two-body transition operator}.
The results of our analysis and a comparison of the different closed-channel realizations are presented in Sec. \ref{subsec: Comparison of the realizations}. 
For the $\ket{bb}$ realization they resemble earlier multichannel studies and allow for an effective field theory description containing a dimer field \cite{Gogolin:2008,Schmidt:2012,Langmack:2018} as we discuss in appendix \ref{app:Second Quantization}. 
In Sec. \ref{subsec: Narrow resonance limit} we analyze the narrow resonance limit of the two realizations.
We then conclude in Sec. \ref{sec:Conclusion and Outlook} with a summary of the most important findings and suggestions for future research.

\section{Three-body bound states in a multichannel setting}
\label{sec:Three-body bound states in a multichannel setting}
For the three-body multichannel model system with separable interactions that we want to focus on in Secs. \ref{sec:Two-channel model system} to \ref{sec:Results}, we choose to work in a momentum space representation. 
This enables us to study Efimov physics by solely considering a one-dimensional integral equation.
Therefore we outline the general multichannel version of the three-body bound state equations for three identical bosons in momentum space here. 
Since we do not yet restrict ourselves to any special kind of model system, the equations presented should apply to any short range interaction including those coupled-channels models, which currently provide the most accurate theoretical description of the interatomic interaction.

We consider identical bosons with pairwise potentials $V_{ij}$ describing the interaction between particles $i$ and $j$.
The three-body bound state equation can then be formulated in the following form \cite{Merkurev:1993_book}
\begin{equation} \label{eq:3bbosons}
\Phi_{ij} = \mathcal{T}_{ij}(E) G_0(E)\left( P_+ + P_- \right) \Phi_{ij} \, ,
\end{equation}
where $\mathcal{T}_{ij}(E)$ is a generalized two-body transition operator, $G_0(E)$ is the free three-body Green's operator, that in the multichannel context also accounts for the asymptotic energies of the channels, and where $P_+$ and $P_-$ denote the two cyclic permutation operators.
The index $(ij)$ specifies a choice of a two particle subsystem formed by particle $i$ and $j$. We introduce the corresponding system of Jacobi momenta
\begin{align}
\bb{p}&=\frac{1}{2}\left(\bb{k}_{j} - \bb{k}_{i} \right)\\
\bb{q}&= \frac{ 1 }{ 3 }( \bb{k}_i + \bb{k}_j )-\frac{ 2 }{ 3 }\bb{k}_k \, ,
\end{align}
where $\bb{k}_i$, $\bb{k}_j$ and $\bb{k}_k$ denote the momentum of particle $i$, $j$ and $k$ respectively, such that the relative momentum between particles $i$ and $j$ is now related to $\mathbf{p}$.
The particular choice of the pair $(ij)$ is abitrary since we are considering identical bosons.
Eq. (\ref{eq:3bbosons}) has a solution only when $E$ is an eigenenergy of the three-body Hamiltonian. The corresponding bound state wave function is then given by $\Psi = (1 + P_+ + P_-)G_0(E)\Phi_{ij}$.

In the following we discuss the operators introduced in Eq. (\ref{eq:3bbosons}) in more detail. 
We start with the analysis
of the generalized two-body transition operator $\mathcal{T}_{ij}(E)$.
Given the interaction $V_{ij}$, we can use the Lippmann-Schwinger equation in order to define $\mathcal{T}_{ij}$
\cite{Merkurev:1993_book}
\begin{equation} \label{eq:Generalized2BT}
\mathcal{T}_{ij}(E) = \left( 1 - V_{ij} G_0(E) \right)^{-1} V_{ij} \, .
\end{equation}
$\mathcal{T}_{ij}(E)$ is related to the two-body $t$ operator $t(z)$ by \cite{Merkurev:1993_book}
\begin{align}
\label{eq:Ttot}
&\langle \mathpzc{C},\mathpzc{c}, \bb{p}, \bb{q}| \mathcal{T}_{ij}(E) | \mathpzc{C}',\mathpzc{c}', \bb{p}', \bb{q}' \rangle \nonumber \\
&= \langle \mathpzc{C}, \bb{p} | t(E - 3\bb{q}^2/4m)| \mathpzc{C}', \bb{p} \rangle \langle \mathpzc{c}, \bb{q} | \mathpzc{c}', \bb{q}' \rangle \, ,
\end{align}
where $\mathpzc{C}$ and $\mathpzc{C}'$ represent any symmetric or antisymmetric combination of the product of internal states of particle $i$ and  $j$ and where $\mathpzc{c}$ and $\mathpzc{c}'$ represent the spin of particle $k$.

Since we consider a three-body system of  identical bosons, where each particle $i$ can occupy several internal spin states labeled by $ \mathpzc{c}_i$,
we can define the permutation operators $P_{\pm}$ as products of permutation operators $P_{\pm}^c$ acting only on coordinates and permutation operators $P_{\pm}^s$ acting only on internal states respectively, such that
\begin{equation}
P_- = P_-^c P_-^s \, \text{ and } P_+ = P_+^c P_+^s \, , \
\end{equation}
where the coordinate permutation operators can be written in integral form as 
\begin{align}
P_-^c &= \int d\bb{q}' d\bb{q} | \bb{q}' + \bb{q}/2, \bb{q}\rangle \langle - \bb{q}'/2 - \bb{q}, \bb{q}'  | \, \\
P_+^c &= \int d\bb{q}' d\bb{q} | - \bb{q}' - \bb{q}/2, \bb{q}\rangle \langle \bb{q}'/2 + \bb{q}, \bb{q}'  | \,  ,
\end{align}
with momentum states normalized according to $\langle \mathbf{p}' | \mathbf{p} \rangle = \delta(\mathbf{p}'-\mathbf{p})$.
The internal state permutation operators can be expressed using a summation over all available internal spin states, such that
\begin{align}
P_-^s &= \sum_{\mathpzc{c}_i \mathpzc{c}_j \mathpzc{c}_k} |\mathpzc{c}_{j},\mathpzc{c}_{k}, \mathpzc{c}_{i} \rangle \langle \mathpzc{c}_{k},\mathpzc{c}_{i},\mathpzc{c}_{j} | \\
P_+^s &= \sum_{\mathpzc{c}_i \mathpzc{c}_j \mathpzc{c}_k} |\mathpzc{c}_{i},\mathpzc{c}_{j}, \mathpzc{c}_{k} \rangle \langle \mathpzc{c}_{k},\mathpzc{c}_{i},\mathpzc{c}_{j} | \, .
\end{align}
The sum of permutation operators can then be written as \cite{Glockle:1983}
\begin{equation}
\label{eq:PpPm}
P_+ + P_- = 2 \left[ \mathcal{P}^S_{ij} P_+ \mathcal{P}^S_{ij} + \mathcal{P}^A_{ij} P_+ \mathcal{P}^A_{ij} \right] \, ,
\end{equation}
where
\begin{equation}
\mathcal{P}^S_{ij} = (1 + P_{ij})/2 \quad \text{and} \quad \mathcal{P}^A_{ij} = (1 - P_{ij})/2
\end{equation}
are the symmetrization and antisymmetrization operators in particles $i$ and $j$ and where $P_{ij}$ is the operator exchanging the particles in the pair $(ij)$.
Eq.~(\ref{eq:PpPm}) then follows from the identity
\begin{equation}
P_- = P_{ij} P_+ P_{ij} \, .
\end{equation}
For identical bosons we need 
$
\Phi_{ij} = P_{ij} \Phi_{ij}
$
such that $\Psi$ is totally symmetric
and therefore Eq. (\ref{eq:3bbosons}) simplifies to
\begin{equation}
\Phi_{ij} = 2 \mathcal{T}_{ij}(E) G_0(E) P_+\Phi_{ij} \, .
\end{equation}
Using this we can give Eq. (\ref{eq:3bbosons}) in $| \mathpzc{C},\mathpzc{c}, \bb{p}, \bb{q}\rangle$ representation, such that
\begin{widetext}
\begin{align} \label{eq:3bbintegralformeval}
& \left\langle \mathpzc{C}, \mathpzc{c} , 
 \bb{p} , 
 \bb{q}
\left|
 \Phi_{ij} 
\right\rangle \right. \nonumber
\\
&=
2 \sum_{\mathpzc{C}',\mathpzc{C}''\mathpzc{c}''} \int d \bb{q}' 
\left\{
\frac{ 
 \left\langle \mathpzc{C}, \bb{p} \left|
t(E - 3\bb{q}^2/4m)\right| \mathpzc{C}', - \bb{q}' - \bb{q}/2 \right\rangle 
}{
E - E(\mathpzc{C}'\beta) - (\bb{q}^2 + \bb{q}'^2 + \bb{q} \bb{q}')/m
} 
\right.
\left. 
\langle \mathpzc{C}',\mathpzc{c} | P_+^s | \mathpzc{C}'',\mathpzc{c}'' \rangle
\left\langle \mathpzc{C}'',\mathpzc{c}'',
\bb{q}'/2 + \bb{q} , \bb{q}'
\left|
 \Phi_{ij}
\right\rangle \right. 
\right\} \, . 
\end{align}
\end{widetext}
Note how the Green's function has evaluated to 
\begin{align}
&G_0(E)|\mathpzc{C}, \mathpzc{c},  \mathbf{p}, \mathbf{q}\rangle \\
&= \frac{| \mathpzc{C}, \mathpzc{c}, \mathbf{p}, \mathbf{q} \rangle}{E-E(\mathpzc{C},\mathpzc{c})-\mathbf{p}^2/m-3\mathbf{q}^2/4m} \, ,
\end{align}
where $E(\mathpzc{C},\mathpzc{c})$ represents the asymptotic energy of the channel $| \mathpzc{C},\mathpzc{c}\rangle $.
In appendix \ref{app:Permutation operators and spin projection} we give more details on how to work out the elements $\langle \mathpzc{C}',\mathpzc{c} | P_+^s | \mathpzc{C}'',\mathpzc{c}'' \rangle$.

\section{Separable model systems}
\label{sec:Two-channel model system}

To explore possible multichannel effects we 
proceed with the analysis of a simple model potential, which has a single separable $s$-wave component
\begin{equation}
\label{eq:seppotential}
\langle \mathpzc{C}, \bb{p} | V| \mathpzc{C}', \bb{p}' \rangle = \langle \mathpzc{C} | v | \mathpzc{C}'\rangle \zeta(p) \zeta^* (p') 
\end{equation}
with form factor $\zeta$ and only symmetric spin combinations $\mathpzc{C}$ and $\mathpzc{C}'$.
In addition we assume just two internal spin states per particle, which we label with $ a $ and $ b $ ($\mathpzc{c}_i, \mathpzc{c}_j, \mathpzc{c}_k \in \{a, b \}$) and assume a difference in channel energy of $\epsilon_{ab}$.
Since the interaction is separable, so is the two-body $t$-operator
\begin{equation}
\langle \mathpzc{C}, \bb{p} | t(z)| \mathpzc{C}', \bb{p}' \rangle = \langle \mathpzc{C} | \tau (z)| \mathpzc{C}'\rangle \zeta(p) \zeta^* (p') \, ,
\end{equation}
that we work out explicitly for a step-function shaped form factor in the following section.
Searching for solutions with zero total angular momentum, we can then make the ansatz $\left\langle \mathpzc{C}, \mathpzc{c} , 
 \bb{p} , 
 \bb{q}
\left|
 \Phi_{ij}
\right\rangle \right.
= \zeta(p) \langle \mathpzc{C}, \mathpzc{c}, 
 q| \phi_{ij} \rangle$.
Evaluating this ansatz in Eq. (\ref{eq:3bbintegralformeval}) and dividing both sides by $\zeta(p)$, we find
\begin{widetext}
\begin{align} \label{eq:3bbintegralformeval2}
&\langle \mathpzc{C}, \mathpzc{c}, 
 q| \phi_{ij} \rangle \nonumber
\\
&=
4 \pi \sum_{\mathpzc{C}',\mathpzc{C}''\mathpzc{c}''} \int_0^\infty q'^2 d q' \int_{-1}^1 d u
\Bigg\{
\frac{ \langle \mathpzc{C}|\tau (E - 3 q^2/4m - E(\mathpzc{c})) |\mathpzc{C}' \rangle  \zeta^* (\sqrt{q'^2 + q^2/4 + q' q u})\zeta (\sqrt{q'^2/4 + q^2 + q' q u}) 
}{
E - E(\mathpzc{C}',\mathpzc{c}) - (q^2 + q'^2 + q q' u)/m
} 
\nonumber\\
& \phantom{= \frac{1}{1}} \qquad \qquad \qquad 
\times \langle \mathpzc{C}' \mathpzc{c} | P_+^s | \mathpzc{C}''\mathpzc{c}'' \rangle
\langle \mathpzc{C}'', \mathpzc{c}'', 
 q'| \phi_{ij} \rangle \Bigg\}   \\
\label{eq:3BBSKform}
&= \sum_{\mathpzc{C}'\mathpzc{c}'} \int_0^\infty q'^2 d q' \langle \mathpzc{C} \mathpzc{c} q| K |\mathpzc{C}' \mathpzc{c}' q' \rangle \langle \mathpzc{C}' \mathpzc{c}' q'| \phi_{ij} \rangle  \, ,
\end{align}
\end{widetext}
where we implicitly defined the operator $K$ in the last line.
We implement Eq. \eqref{eq:3bbintegralformeval2} numerically by replacing the $q'$ integration by a summation over a finite $q'$-grid.

Applying the relations as presented in appendix \ref{app:Permutation operators and spin projection} 
in order to analyze the elements 
$ \langle \mathpzc{C}' \mathpzc{c} |P_+^s|\mathpzc{C}'' \mathpzc{c}'' \rangle $ and dropping the index in $\phi_{ij}$,
 Eq.~(\ref{eq:3BBSKform}) can be rewritten into the following matrix form

\begin{widetext}
{
\begin{align}
\label{eq:3bbmatrixform}
&\left[
\begin{array}{c}
\langle \underline{aa}a| \phi \rangle \\
\langle \underline{ab}a| \phi \rangle \\
\langle \underline{bb}a| \phi \rangle \\
\langle \underline{aa}b| \phi \rangle \\
\langle \underline{ab}b| \phi \rangle \\
\langle \underline{bb}b| \phi \rangle \\
\end{array}
\right] \nonumber \\
&=
\left[
\begin{array}{cccccc}
 \langle \underline{aa}a|K|\underline{aa}a \rangle 
 & \langle \underline{aa}a|K|\underline{ab}a \rangle  & 0 & \langle \underline{aa}a|K|\underline{aa}b \rangle  &\langle \underline{aa}a|K|\underline{ab}b \rangle  & 0 \\
 \langle \underline{ab}a|K|\underline{aa}a \rangle  & \langle \underline{ab}a|K|\underline{ab}a \rangle & 0 & \langle \underline{ab}a|K|\underline{aa}b \rangle & \langle \underline{ab}a|K|\underline{ab}b \rangle & 0 \\
 \langle \underline{bb}a|K|\underline{aa}a \rangle & \langle \underline{bb}a|K|\underline{ab}a \rangle & 0 & \langle \underline{bb}a|K|\underline{aa}b \rangle & \langle \underline{bb}a|K|\underline{ab}b \rangle & 0 \\
 0 & \langle \underline{aa}b|K|\underline{ab}a \rangle & \langle \underline{aa}b|K|\underline{bb}a \rangle & 0 & \langle \underline{aa}b|K|\underline{ab}b \rangle & \langle \underline{aa}b|K|\underline{bb}b \rangle \\
 0 & \langle \underline{ab}b|K|\underline{ab}a \rangle & \langle \underline{ab}b|K|\underline{bb}a \rangle & 0 & \langle \underline{ab}b|K|\underline{ab}b \rangle & \langle \underline{ab}b|K|\underline{bb}b \rangle \\
 0 & \langle \underline{bb}b|K|\underline{ab}a \rangle & \langle \underline{bb}b|K|\underline{bb}a \rangle & 0 & \langle \underline{bb}b|K|\underline{ab}b \rangle & \langle \underline{bb}b|K|\underline{bb}b \rangle \\
\end{array}
\right]
\left[
\begin{array}{c}
\langle \underline{aa}a| \phi \rangle \\
\langle \underline{ab}a| \phi \rangle \\
\langle \underline{bb}a| \phi \rangle \\
\langle \underline{aa}b| \phi \rangle \\
\langle \underline{ab}b| \phi \rangle \\
\langle \underline{bb}b| \phi \rangle \\
\end{array}
\right] \, ,
\end{align}}
\end{widetext}
where we write out the symmetric spin states in the pair ($ij$) explicitly as $\mathpzc{C}\in\{ \underline{aa}, \underline{ab}, \underline{bb} \} $.
This equation indicates, that even a model with a single $s$-wave component and two internal states, yields a system with six coupled channels for the Faddeev component $\phi$. 

By setting $\langle \mathpzc{C}'|v|\mathpzc{C}\rangle = 0$ when $\mathpzc{C}$ or $\mathpzc{C}'$ equals $ \underline{ab}$, or alternatively setting $\langle \mathpzc{C}'|v|\mathpzc{C}\rangle = 0$ when $\mathpzc{C}$ or $\mathpzc{C}'$ equals $\underline{bb}$ we can study scenarios where the closed symmetric spin channel $Q$ is either of the form $|\mathpzc{c} \mathpzc{c} \rangle $ or $ | \mathpzc{c} \mathpzc{c}' \rangle_S = (| \mathpzc{c} \mathpzc{c}' \rangle + | \mathpzc{c}' \mathpzc{c} \rangle) / \sqrt{2} $, respectively.
Thereby we reduce to a two-channel model on the two-body level with open channel $P =\underline{aa}$ in both of these cases. 
As we will see in the following section $\langle \mathpzc{C}'|\tau(z)|\mathpzc{C}\rangle = 0$ whenever $\langle \mathpzc{C}'|v|\mathpzc{C}\rangle = 0$.
Therefore some of the elements $\langle \mathpzc{C}' \mathpzc{c}'|K| \mathpzc{C} \mathpzc{c} \rangle $ in Eq. (\ref{eq:3bbintegralformeval2}) evaluate to zero and the three-body equations can be simplified.

Considering Eq. (\ref{eq:3bbmatrixform}) for the situation with closed channel $Q = \underline{bb}$, we find that the three-body equations involving the open channel $\underline{aa}a$ can be reduced to
\begin{align}
\left[
\begin{array}{c}
\langle \underline{aa}a| \phi \rangle \\
\langle \underline{bb}a| \phi \rangle \\
\end{array}
\right]
=
\left[
\begin{array}{cc}
 \langle \underline{aa}a|K|\underline{aa}a \rangle  & 0  \\
 \langle \underline{bb}a|K|\underline{aa}a \rangle & 0 \\
\end{array}
\right]
\left[
\begin{array}{c}
\langle \underline{aa}a| \phi \rangle \\
\langle \underline{bb}a| \phi \rangle \\
\end{array}
\right] \, .
\end{align}
The solution is thus solely determined by the open channel part
\begin{equation}\label{eq:singlechannelbb}
\langle \underline{aa}a| \phi \rangle = \langle \underline{aa}a |K| \underline{aa}a \rangle \langle \underline{aa}a| \phi \rangle \, .
\end{equation}
Since just the open-channel component of the $t$-operator is needed in this equation, we can later on use the Feshbach formalism \cite{Feshbach92}, which we generalize in appendix \ref{app:The transition matrix for the bb-channel configuration} to an off shell version, to construct an approximate model system, in which the closed-channel is modelled by a separable energy-dependent interaction term added to the open channel.
We analyze this model and its narrow resonance limit in more detail in Sec. \ref{sec:Results}.
We note that this model system is similar to the ones discussed in effective field theory \cite{Gogolin:2008,Schmidt:2012,Langmack:2018}
and the ones considered for the narrow resonance limit \cite{Petrov:2004,Gogolin:2008}.
In appendix \ref{app:Second Quantization} we show that the $Q=\underline{bb}$ model can be approximated by the effective field theory models used in \cite{Gogolin:2008,Schmidt:2012,Langmack:2018}.

If we alternatively fix the closed channel $Q$ to correspond to $\underline{ab}$, we find that Eq. \eqref{eq:3bbmatrixform} can be expressed as
\begin{widetext}
{
\begin{align}
\label{eq:3BBSabmatrix}
\left[
\begin{array}{c}
\langle \underline{aa}a| \phi \rangle \\
\langle \underline{ab}a| \phi \rangle \\
\langle \underline{aa}b| \phi \rangle \\
\langle \underline{ab}b| \phi \rangle \\
\end{array}
\right]
=
\left[
\begin{array}{cccc}
 \langle \underline{aa}a|K|\underline{aa}a \rangle  & \langle \underline{aa}a|K|\underline{ab}a \rangle   & \langle \underline{aa}a|K|\underline{aa}b \rangle  & 0   \\
 \langle \underline{ab}a|K|\underline{aa}a \rangle  & \langle \underline{ab}a|K|\underline{ab}a \rangle  & \langle \underline{ab}a|K|\underline{aa}b \rangle & 0  \\
 0 & \langle \underline{aa}b|K|\underline{ab}a \rangle & 0 & \langle \underline{aa}b|K|\underline{ab}b \rangle \\
 0 & \langle \underline{ab}b|K|\underline{ab}a \rangle  & 0 & \langle \underline{ab}b|K|\underline{ab}b \rangle  \\
\end{array}
\right]
\left[
\begin{array}{c}
\langle \underline{aa}a| \phi \rangle \\
\langle \underline{ab}a| \phi \rangle \\
\langle \underline{aa}b| \phi \rangle \\
\langle \underline{ab}b| \phi \rangle \\
\end{array}
\right] \, .
\end{align}}
\end{widetext}
Contrary to the realization where $Q = \underline{bb}$, this equation indicates that the realization $Q = \underline{ab}$ results in coupling terms to the closed channels. 
Consequently it is no longer possible to describe the model in terms of the Feshbach formalism. 
As such, the system does no longer resemble the ones discussed in effective field theory \cite{Gogolin:2008,Schmidt:2012,Langmack:2018} and requires a careful analysis in terms of all four coupled-channels as presented in Eq. (\ref{eq:3BBSabmatrix}). 
The results of this analysis will be presented in Sec. \ref{subsec: Comparison of the realizations}.

\section{Two-body transition operator}
\label{sec:Two-body transition operator}

In the following we derive the two-body $t$-operator for the separable system explicitly.
We consider the cases of open channel $P = \underline{aa} $ and closed channel $Q = \underline{bb}$ as well as open channel $P = \underline{aa} $ and closed channel $Q = \underline{ab}$.
Since we can reduce to a two-body two-channel system in both cases, the $t$-operators are identical, when the difference $\epsilon$ in closed and open channel energy and the coupling strengths $\langle P | v | P \rangle, \,\langle Q | v | Q \rangle, \,\langle Q | v | P \rangle = \langle Q | v | P \rangle^*$ are the same.
We get $\epsilon = 2 \epsilon_{ab}$ for $Q=\underline{bb}$ and $\epsilon = \epsilon_{ab}$ for $Q=\underline{ab}$. 
According to Eq. (\ref{eq:seppotential}) we have a two-body interaction of the form
\begin{equation}
V
= 
\left(
\begin{matrix}
\bar{v}_{PP} & \bar{v}_{PQ} \\
\bar{v}_{QP} & \bar{v}_{QQ}
\end{matrix}
\right) \frac{
|\zeta \rangle \langle \zeta|}{m \Lambda} = [v] |\zeta \rangle \langle \zeta |
\, ,
\end{equation}
with potential interaction and coupling strength parameters $\bar{v}_{PP}, \, \bar{v}_{QQ}, \, \bar{v}_{PQ} = \bar{v}_{QP} \in \mathbb{R}$.
We define the form factor $\zeta$ as
\begin{equation}
\langle \bb{p} | \zeta \rangle
= \zeta(p) = \begin{cases}
 1\, , p <\Lambda\\
 0 \, , p \geq \Lambda
 \end{cases} 
 \, ,
\end{equation}
where $\Lambda$ is a momentum cut-off scale.
For such an interaction, the two-body transition operator can be obtained analytically in a straightforward fashion.
Resembling Eq.~(\ref{eq:Generalized2BT}) the two-body operator is defined as
\begin{equation}
t(z)
=
\left(
1 - V g_0(z)
\right)^{-1} V \, ,
\end{equation}
with $g_0$ the free Green's function of the two-body system. 
Since we are considering the interaction between just two channels, we can fix the zero energy to equal the asymptotic energy of the open channel $P$. 
By doing so, the asymptotic energy of the closed-channel $Q$ reduces to the energy difference $ \epsilon$ between the open and closed channel. 
Consequently, the free Green's operator of the two-body system $g_0(z)$ can be expressed as
\begin{align}
&\langle \bb{p}| g_0(z) | \bb{p}' \rangle \\
&= 
\left(
\begin{matrix}
\left(z 
- p^2 / m  \right)^{-1} & 0 \\
0 & \left(z - \epsilon
-  p^2 / m \right)^{-1} 
\end{matrix}
\right) \langle \bb{p} | \bb{p}' \rangle \, . \nonumber
\end{align}
Next, the separable interactions allow us to express the $t$-operator in the following separable form
\begin{equation}
t(z) = 
\left(
\begin{matrix}
\tau_{PP}(z) & \tau_{PQ}(z) \\
\tau_{QP}(z) & \tau_{QQ}(z)
\end{matrix}
\right)
|\zeta \rangle \langle \zeta | 
\, ,
\end{equation}
where the energy dependent terms $t_{\mathpzc{C} \mathpzc{C}'}(z)$ can be computed explicitly from
\begin{equation} \label{eq:toperator2channel}
\left(
\begin{matrix}
\tau_{PP}(z) & \tau_{PQ}(z) \\
\tau_{QP}(z) & \tau_{QQ}(z)
\end{matrix}
\right)
=
\left(
1 - [v] \langle \zeta | g_0(z)| \zeta \rangle
\right)^{-1} [v] \, .
\end{equation}
For step function shaped form factors considered in this section, the previous equation can be solved analytically
using the identity
\begin{align}
&\int_0^\Lambda dp \frac{p^2}{p_z^2 - p^2 + i 0} \nonumber \\
&= - \Lambda + p_z \mathrm{arctanh}\left(\frac{\Lambda}{p_z}\right) 
\end{align}
to solve for $\langle \zeta | g_0(z)| \zeta \rangle$.

Here we would like to point out that it is possible to generalize this method to any finite number of channels and form factors.
Furthermore, the model can be adjusted to match the low energy scattering properties of a given system.
The scattering length is then given by 
\begin{equation}
a = 2 \pi^2 m \hbar \tau_{PP}(0) \zeta^*(0) \zeta(0)
\end{equation}
and depends on the parameters $\bar{v}_{PP}$, $\bar{v}_{QQ}$, $\bar{v}_{PQ} = \bar{v}_{QP}$ and $\epsilon$, so that we have $a = a(\bar{v}_{PP},\bar{v}_{QQ},\bar{v}_{PQ},\epsilon)$.
We then define the background scattering length by 
\begin{equation}
a_\mathrm{bg}(\bar{v}_{PP},\bar{v}_{QQ},\bar{v}_{PQ}) = \lim_{\epsilon \rightarrow \infty} a(\bar{v}_{PP},\bar{v}_{QQ},\bar{v}_{PQ},\epsilon) \, , 
\end{equation}
the resonance energy $\epsilon_0$ by 
\begin{equation}
1/a(\bar{v}_{PP},\bar{v}_{QQ},\bar{v}_{PQ},\epsilon_0(\bar{v}_{PP},\bar{v}_{QQ},\bar{v}_{PQ}))=0
\end{equation} 
and the resonance width parameter
\begin{equation}
r^* = \left. \partial_\epsilon \left( \frac{m}{\hbar^2 a} \right) \right|_{\epsilon = \epsilon_0} \, .
\end{equation}
With those definitions we can map any given set of $(a_\mathrm{bg},r^*, \epsilon_0)$ to a set of system parameters $(\bar{v}_{PP},\bar{v}_{QQ},\bar{v}_{PQ})$.

To analyze the narrow resonance limit of $\tau_{PP}(z)$ we first use the Feshbach formalism outlined in appendix \ref{app:The transition matrix for the bb-channel configuration} to approximate the system.
We recognize that the transition matrix is the only operator in the three-body equation that depends on the form of the two-body interactions. Following the procedure as outlined in appendix \ref{app:The transition matrix for the bb-channel configuration}, we find that the transition matrix element $\tau_{PP}$ as introduced in Eq. (\ref{eq:toperator2channel}) reduces to the following simple form in the narrow resonance limit 
\begin{align}
\label{eq:tppNarrowLimit1}
\tilde{\tau}_{PP} &\underset{\phantom{r^* \rightarrow \infty}}{\equiv} \frac{\tau_{PP}(z)}{r^*/m \hbar } \\
&\underset{r^* \rightarrow \infty}{=} 
\frac{1}{2 \pi^2 |\zeta(0)|^2 } \times \frac{1}{\tilde{z} +\tilde{a}^{-1}-\sqrt{-\tilde{z}}} \, , \label{eq:tppNarrowLimit}
\end{align}
where we have used system parameters in units of $r^*$, such that $\tilde{t}_{PP} = t_{PP}/( r^*/m \hbar)$, and where we have introduced the dimensionless scattering length $\tilde{a}= a / r^*$ and 
energy $\tilde{z} = z \, m \, r^{*2}/ \hbar^2$. 
The above expression is valid for arbitrary form factors $\zeta$ and leads to a narrow resonance limit of the Efimov spectrum which we discuss in detail in section \ref{subsec: Narrow resonance limit}.
From that we conclude that the above limit also holds in the general setting with non separable interaction potentials. 
The $t$-operator for those general potentials can be expanded in separable terms, with a single separable term representing the resonant component \cite{Mestrom:2019,Secker:2020e,Secker:2020b}.
Only the open-open component of the resonant term will approach the limit in Eq.~(\ref{eq:tppNarrowLimit}).
All other terms in the open-open component are  finite even on resonance in units related to the range of the interaction and therefore vanish according to Eq.~(\ref{eq:tppNarrowLimit1}) in units of the resonance width parameter $r^*$ when taking it to infinity.
\\

\section{Results}
\label{sec:Results}
We study the dissociation 
scattering lengths $a_-^{(n)}$ as well as the binding energies of the three deepest trimer states for varying values of the resonance width parameter $r^*$.
To completely determine the system we fix the threshold difference on resonance $\epsilon_0$ and the background scattering length $a_{bg}$, such that we can map the scattering length $a$ to the threshold difference $\epsilon$ between the open and closed channel.
In the following we denote quantities made dimensionless in units of the cut-off scale $\Lambda$ with a bar.
This means that all lengths are given in multiples of $\hbar / \Lambda $ and all energies are given in multiples of $ \Lambda^2 / m$.
We choose $\bar{a}_{bg} = -0.2$ to have no additional bound background dimer states in our model and set the resonance position to a value of $\bar{\epsilon}_0 = 1.5$ to just have a single closed channel trimer for $Q = \underline{ab}$ as will be discussed below in more detail.

\subsection{Comparison of the $Q = \underline{bb}$ and $Q = \underline{ab}$ multichannel realizations}
\label{subsec: Comparison of the realizations}

\begin{figure*}[t]
\centering
\includegraphics[width= \textwidth]{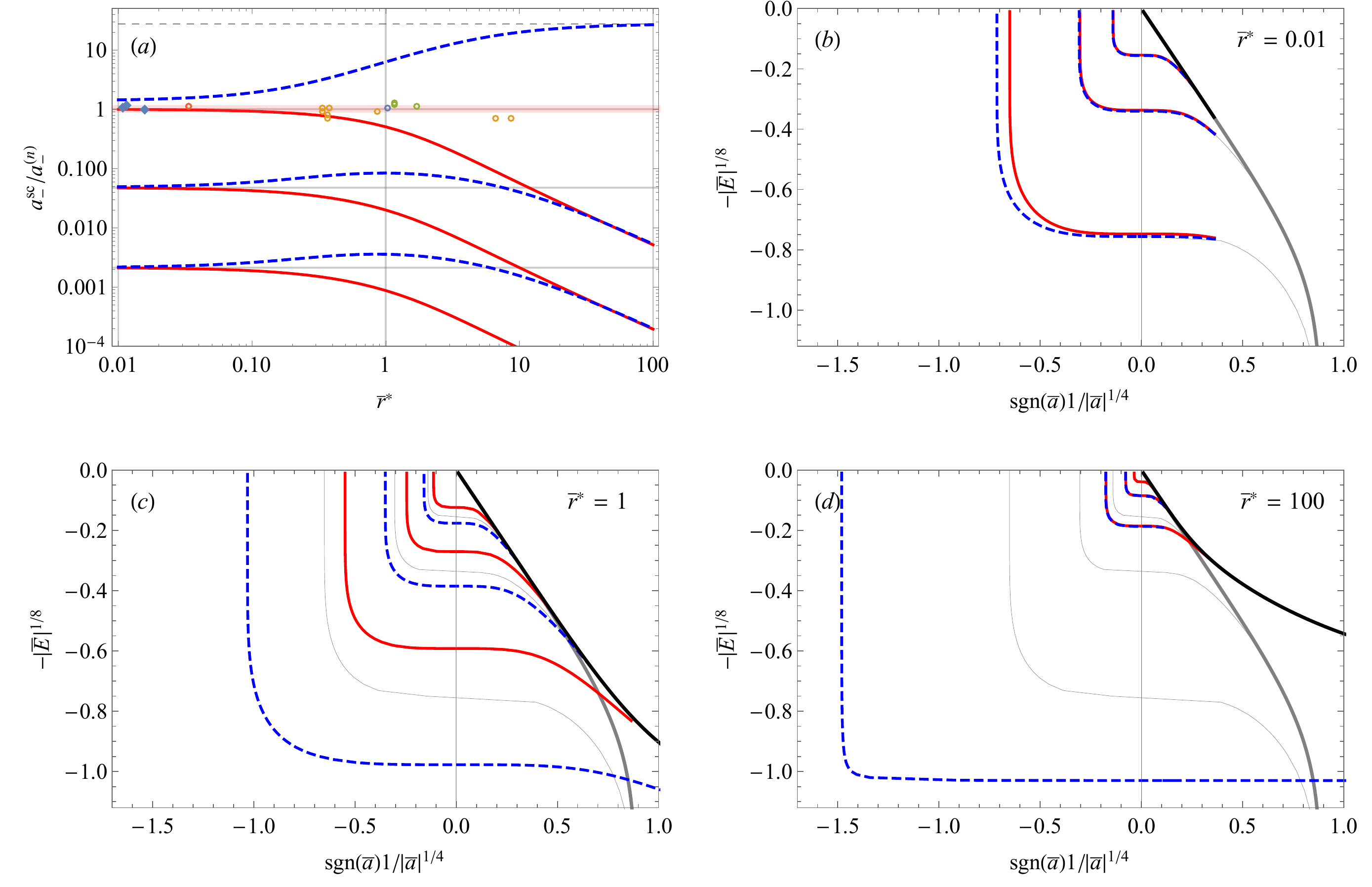}
\caption{We compare the three-body spectra of the systems we have analyzed. 
The red solid lines correspond to the $Q=\underline{bb}$ realization, whereas the blue dashed lines correspond to the $Q = \underline{ab}$ realization of the system.
In both cases the resonance positions are fixed at a threshold difference of $\bar{\epsilon}_0 = 1.5$ and the background scattering length is fixed to $\bar{a}_{bg} = -0.2$.
In (a) we show the inverse dissociation scattering lengths $1 / a_-^{(n)}$ of the three most deeply bound trimer states rescaled with the lowest (in absolute value) single-channel dissociation scattering length $a_-^{sc}$ as a function of the resonance width $\bar{r}^*$. 
The open circles correspond to experimental data \cite{Gross:2011, Dyke:2013, Chapurin:2019, Roy:2013, Wild:2012,Berninger:2011,Kraemer:2006,Wang:2014}, where we rescaled $a_-^{(0)}$ by the universal value of $-9.73 \, r_\mathrm{vdW}$ \cite{Wang:2012} and set $\bar{r}^* = r^* / r_\mathrm{vdW}$. 
The red band indicates the van der Waals universal region up to $\pm 15 \%$. 
The diamonds correspond to experimental data \cite{Berninger:2011} related to very small values of $r^*$ and have been shifted by $\bar{r}^* = 0.01$ to fit on the plot.
The thin black dashed line indicates $a_-^\mathrm{sc}/a_\mathrm{bg}$, while the light gray horizontal lines correspond to the trimer positions of the single-channel system.
The vertical gray lines indicate the resonance widths at which we obtained full Efimov spectra shown in (b) - (d) for $\bar{r}^* = 0.01, 1, 100$ respectively.
In Figs. (b) - (d) we show the binding energies of the lowest three trimer states (red full and blue dashed lines) as a function of the inverse scattering length $1/\bar{a}$. 
For better visibility both axes are rescaled as indicated by the axes labels.
The thick black line indicates the dimer binding energy which agrees for both realizations since the $t$-operator is identical.
The thin gray lines correspond to the trimer spectrum of the single-channel model, while the thick gray line corresponds to the single-channel dimer binding energy.
In Fig.~\ref{fig:fig1}(b) the trimer and dimer lines stop at some positive inverse scattering length close to $1/\bar{a}^{1/4} = 0.4$, since the threshold difference related to this value of the scattering length is zero and we enter a regime irrelevant for this investigation beyond this point.}
\label{fig:fig1}
\end{figure*}

We start by comparing our results for $a_-^{(n)}$ in the different closed channel realizations $Q = \underline{bb}$ and $Q = \underline{ab}$ for changing values of $r^*$. 
Our results are summarized in Fig. \ref{fig:fig1}(a).
Here we also compare our results to the single-channel result corresponding to the interaction \cite{Naidon:2017,Colussi:2018}
\begin{equation}
V^{\mathrm{sc}} = \frac{\bar{a}}{2 \pi^2 - 4 \pi \bar{a}} \frac{| \zeta \rangle \langle \zeta |}{m \Lambda} \, .
\end{equation}
We find that in the broad resonance limit ($\bar{r}^* \rightarrow 0$) we can recover the single-channel result for all $a_-^{(n)}$ except for the dissociation scattering length $a_-^{(0)}$
related to the lowest Efimov trimer state in the $Q = \underline{ab}$ realization, which still approaches a limiting value but is lowered in absolute value. 

Our finding that a limiting result is approached for broad resonances is in agreement with earlier studies in effective field theory \cite{Schmidt:2012, Langmack:2018} and is also in line with the results of multichannel hyperradial calculations using van der Waals interaction potentials \cite{Wang:2014}.
Also the experimental results for atomic systems indicated by the blue diamonds in Fig.~\ref{fig:fig1}(a) confirm this behavior \cite{Berninger:2011}.
When $\bar{r}^*$ is increased we find that $|a_-^{(n)}|$ increases for the $Q = \underline{bb}$ realization in agreement with \cite{Schmidt:2012,Langmack:2018}.
However, for $Q = \underline{ab}$ we find the opposite behavior and $|a_-^{(n)}|$ decreases with increasing $\bar{r}^*$, for moderate values of $\bar{r}^* < 0.3$.
This behavior is more in line with the results of \cite{Sorensen:2012}, where a lowering in $|a_-^{(0)}|$ was observed when $\bar{r}^*$ is increased.
The value of $|a_-^{(0)}|$ corresponding to the lowest trimer state in the $Q = \underline{ab}$ realization even keeps decreasing when $\bar{r}^*$ reaches large values and converges to $\lim_{\bar{r}^* \rightarrow \infty} a_-^{(0)}  = a_{bg}$.
This indicates that the lowest trimer state in the $Q = \underline{ab}$ realization is in this limit no longer related to the Efimov spectrum close to the resonance and has in fact purely closed-channel character, as will be discussed in the following section.
Note that due to 
\begin{align}
a (\epsilon) & \approx a_\mathrm{bg} - \frac{\hbar^2/ m r^*}{ \epsilon_0 - \epsilon}
\end{align}
all fixed $\epsilon$ will be mapped  to $a_{bg}$ in the narrow resonance limit.
In this limit ($\bar{r}^* \rightarrow \infty$) we also find dissociation scattering lengths related to Efimov states that scale linearly with $r^*$ for both realizations $Q = \underline{bb}$ and $Q = \underline{ab}$.
We discuss this limit in more detail in section \ref{subsec: Narrow resonance limit}.

In Fig.~\ref{fig:fig1}(b) - (d) we show some trimer spectra for increasing values of $r^*$.
These spectra may be most easily understood starting from the narrow resonance spectrum given in Fig. \ref{fig:fig1}(d).
In Fig. \ref{fig:fig1}(d) we identify the closed-channel trimer as the one that is most deeply bound in the
plot in the $Q = \underline{ab}$ realization.
Close to the point where the dimer state merges with the three-body continuum we find a shrunken version of a universal Efimov trimer spectrum for both realizations $Q = \underline{bb}$ and $Q = \underline{ab}$.
The thin gray lines give the single-channel Efimov spectrum for comparison.
As mentioned earlier the size of the Efimov trimer spectrum is set by $r^*$, which is the dominating length scale in the narrow resonance limit.
Therefore the size of the Efimov spectrum increases when $r^*$ decreases as can be seen from Fig. \ref{fig:fig1}(c).
In this intermediate resonance width regime where $\bar{r}^* \sim 1$ deviations between the different realizations $Q = \underline{bb}$ and $Q = \underline{ab}$ get pronounced, since the excited trimer states in the $Q = \underline{ab}$ realization start to couple to and get repelled by the lowest closed-channel trimer state.
This causes the first and second excited trimer states to be shifted to higher energies as compared to the energies of the two lowest Efimov trimer states in the $Q = \underline{bb}$ realization.
Due to the coupling to the open-channel ($\underline{aa}a$) Efimov trimers the closed-channel ($\underline{ab}a$ and $\underline{aa}b$) trimer acquires an open-channel component. 
We note that in the $Q = \underline{bb}$ realization it is not possible to couple to closed-channel trimer states, as is indicated by Eq. (\ref{eq:singlechannelbb}).
For even broader resonances our results are shown in Fig. \ref{fig:fig1}(b).
There the lowest trimer state in the $Q = \underline{ab}$ realization adopts Efimov character, while the first and second excited trimer energies are matching with the ones of the $Q = \underline{bb}$ realization as well as with the single-channel result.

We also compare to experimental results in Fig. \ref{fig:fig1}(a). 
We find that almost all experimentally measured values of $a_-^{(0)}$ lie between the predictions of the $Q = \underline{bb}$ and $Q = \underline{ab}$ realizations, when rescaled according to the description in the caption of Fig.~\ref{fig:fig1}(a). 
This is promising since interpolating between the $Q = \underline{bb}$ and $Q = \underline{ab}$ interaction potentials while also properly adjusting $\epsilon_{ab}$ provides us with a continuous mapping between the two limiting realizations we study.
Therefore also the spectra should be continuously deformed into each other covering part of the area between the models, which contains the experimental values.
This indicates that in the realistic situation a model including both realizations $Q = \underline{bb}$ and $Q = \underline{ab}$ needs to be applied to represent the atomic spin structure correctly.
We note that in a realistic system usually both $| \mathpzc{c} \mathpzc{c} \rangle $ and $| \mathpzc{c} \mathpzc{c}' \rangle_S$ type channels are involved in the multichannel interaction \cite{Secker:2020b}.

\subsection{Narrow resonance limit}
\label{subsec: Narrow resonance limit}

For the $Q = \underline{bb}$ realization we can immediately see from Eq.~(\ref{eq:singlechannelbb}) that the trimer energies depend only on the three-body open channel component $\langle \underline{aa}a | \phi \rangle$.
For the $Q = \underline{ab}$ realization on the other hand the coupling terms $\langle \underline{aa}a |K| \underline{ab}a \rangle$ and $\langle \underline{aa}a |K| \underline{aa}b \rangle$, prevent this.
However, these coupling terms vanish in the narrow resonance limit when expressed in units related to the width parameter $r^*$,
because the separation in threshold energy is $ \tilde{E}(\underline{ab}a) - \tilde{E}(\underline{aa}a) = \tilde{\epsilon} \propto \bar{r}^{*2}$.
Therefore $G_0$ has a suppressing effect $\propto 1/\bar{r}^{*2}$, which cancels the leading order diverging behavior of $\tilde{\tau}_{PQ}(z)\propto \sqrt{\bar{r}^*}$ in the coupling terms $\langle \underline{aa}a |K| \underline{ab}a \rangle$ and $\langle \underline{aa}a |K| \underline{aa}b \rangle$.
In conclusion we find that in the narrow resonance limit Eq. (\ref{eq:singlechannelbb}) can be used to solve for three-body bound state energies in both realizations $Q = \underline{bb}$ and $Q = \underline{ab}$.
Since the scaling behavior of $\tilde{\tau}_{PQ}(z)\propto \sqrt{\bar{r}^*}$ holds in general, the above reasoning is also true for realistic interactions including the full spin structure of the three-atom system.
We thus conclude that the above limit holds in the narrow resonance limit for any multichannel interaction potential.

By applying the reduction given in Eq. (\ref{eq:tppNarrowLimit}) and by changing to the scaled momenta $\tilde{q} = q r^* / \hbar$, we find that $q = \hbar\tilde{q}/ r^* $ approaches a value of zero in the low-energy and narrow resonance limit.
We can therefore simplify the expression of the integral kernel as presented in Eq. (\ref{eq:3bbintegralformeval2}) by replacing the argument in the form-factors $\zeta$ by the value at zero momentum $\zeta(0)$. 
The $\zeta(0)$-terms cancel with the ones contained in $\tilde{\tau}_{PP}(z)$ (compare Eq. (\ref{eq:tppNarrowLimit})).
Consequently, the three-body wave function of any model with separable interactions can be expressed as 
\begin{align}
\langle \underline{aa}a, \tilde{q}|\phi \rangle & \underset{r^* \rightarrow \infty}{\approx}
4\pi \int d \tilde{q}' \frac{\tilde{q}' \langle \underline{aa}a,\tilde{q}'|\phi \rangle}{2 \pi^2 \tilde{q} (-\tilde{q}_z^2 +\tilde{a}^{-1}-\tilde{q}_z)}  \notag \\[3pt] 
& \phantom{\underset{r^* \rightarrow \infty}{\approx}} \times \text{log}\left(\frac{-\tilde{E}+\tilde{q}'^2+\tilde{q}^2-\tilde{q}'\tilde{q}}{-\tilde{E}+\tilde{q}'^2+\tilde{q}^2+\tilde{q}'\tilde{q}}\right),
\label{eq:narrow3Blimit}
\end{align}
where the absence of form factors has allowed us to carry out the angular integration explicitly and where $\tilde{q}_z = \sqrt{3 \tilde{q}^2 / 4 - \tilde{E}}$. 

\begin{figure}[t]
\centering
\includegraphics[width= \columnwidth]{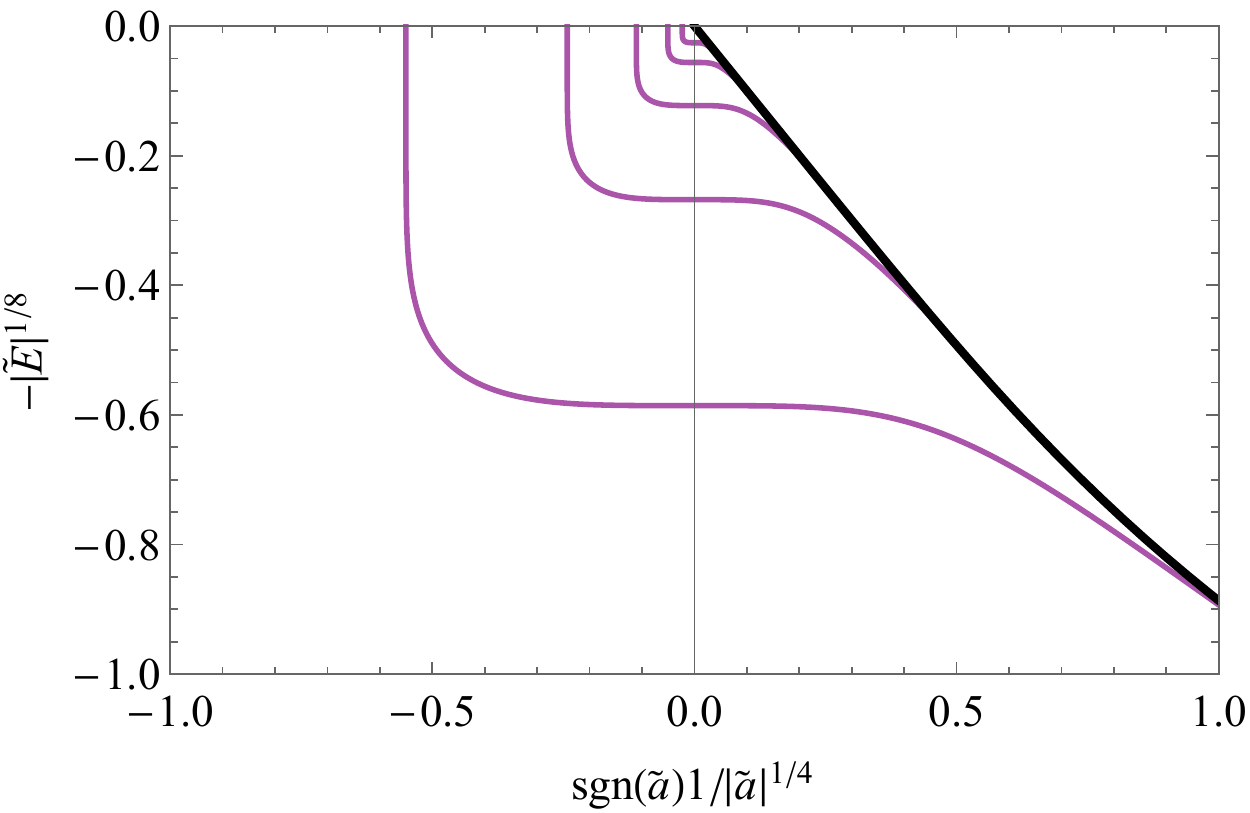}
\caption{Efimov spectrum in the narrow resonance limit showing the first five Efimov trimers (purple) and the ground-state dimer (thick black) in dimensionless units of $r^*$.}
\label{fig:EfimovRstarDimlessArticle}
\end{figure}

As we have found a straightforward expression Eq. (\ref{eq:narrow3Blimit}) for the three-body bound state equation in the narrow resonance limit, we can proceed with the computation of the dimensionless Efimov spectrum. 
In Fig. \ref{fig:EfimovRstarDimlessArticle} we present our results for the Efimov spectrum in the narrow resonance limit.
In addition we extract the dissociation scattering length $a_{-}^{(n)}$ up to $n=3$ as well as the wave number $\kappa_*^{(n)}$ related to the trimer binding energy on resonance. 
Our results are collected in Tab. \ref{tab:EfimovTable}. 

\begin{table}[H]
 \caption{Values of $\tilde{a}_-^{(n)} e^{-n\pi/s_0}$ as well as $a_-^{(n+1)}/a_-^{(n)}$ and $\kappa_*^{(n)}a_-^{(n)}$ for the four lowest Efimov trimers.}
    \label{tab:EfimovTable}
  \begin{center}
    \begin{tabularx}{\columnwidth}{ c| >{\centering \arraybackslash}X>{\centering \arraybackslash}X>{\centering \arraybackslash}X } 
    \hline 
    \hline
      $n$ & $\tilde{a}_-^{(n)} e^{-n\pi/s_0}$  & $a_-^{(n+1)}/a_-^{(n)}$ & $\kappa_*^{(n)}a_-^{(n)}$ \\
      \hline 
	  $0$ & $−10.90$ & $26.48$ & $1.28$ \\
      $1$ & $−12.72$ & $22.98$ & $1.49$ \\
      $2$ & $−12.88$ & $22.71$ & $1.51$ \\
      $3$ & $−12.90$ & $22.70$ & $1.51$ \\
      \hline  
      \hline
    \end{tabularx}
  \end{center}
\end{table}
In agreement with Refs. \cite{Gogolin:2008,Nishida:2012,PhysRevA.84.062704,Endo:2016,Naidon:2017}, we find in the narrow resonance limit that the dissociation scattering length of the ground state $a_-^{(0)}$ scales as $a_-^{(0)} / r^* \simeq -10.90216 $, whereas highly excited trimer states $(n \gg 1)$ approach a scaling $a_-^{(n)}e^{-n \pi/s_0}/r^* \simeq -12.9$. 
In addition, we recognize the universal scaling laws $a_-^{(n+1)}/a_-^{(n)} \approx -22.7$ and $\kappa_*^{(n)}a_-^{(n)} \approx 1.51$ which are typical for Efimov spectra.  

Having analyzed the trimer spectrum in the open three-body channel $\underline{aa}a$ in the narrow resonance limit, we are left with the narrow resonance limit analysis of the closed-channel trimer spectrum for the $Q = \underline{ab}$ realization.
In the narrow resonance limit ($r^*\rightarrow \infty$) we find that $\bar{\tau}_{PQ}(E - 3q^2/4m) \rightarrow 0 $ for all $E$ considered.
It follows that the coupling terms $\langle \underline{aa}a |K| \underline{ab}a \rangle$, $\langle \underline{aa}a |K| \underline{aa}b \rangle$, $\langle \underline{ab}a |K| \underline{aa}a \rangle$, $\langle \underline{aa}b |K| \underline{ab}b \rangle$ and $\langle \underline{ab}b |K| \underline{ab}a \rangle$ vanish.
We can therefore analyze the closed-channel components $\underline{ab}a$ and $\underline{aa}b$ separately, which leads to the following system of equations
\begin{align}
&\left[
\begin{array}{c}
\langle \underline{ab}a| \phi \rangle \\
\langle \underline{aa}b| \phi \rangle \\
\end{array}
\right] \nonumber \\
&=
\left[
\begin{array}{cc}
 \langle \underline{ab}a|K|\underline{ab}a \rangle  & \langle 
\underline{ab}a|K|\underline{aa}b \rangle  \\
 \langle \underline{aa}b|K|\underline{ab}a \rangle & 0  \\
\end{array}
\right]
\left[
\begin{array}{c}
\langle \underline{ab}a| \phi \rangle \\
\langle \underline{aa}b| \phi \rangle \\
\end{array}
\right] \, ,
\end{align}
The system consists of two particles in spin state $|a \rangle$ and one particle in spin state $|b \rangle$ and is therefore equivalent to a system of two identical bosons (B) and a distinguishable particle (X) (see also appendix \ref{app:Second Quantization}).
The Efimov scaling laws on resonance for such a BBX system are determined by $s_0 \approx 0.41370$ \cite{Naidon:2017,Colussi:2016}.
We find our system in agreement with those results, when setting $a_\mathrm{bg} = 0$ with $s_0 = 0.415$ that we determined from the scaling of the ground to first excited Efimov trimer energy on resonance.

\begin{figure}[H]
\centering
\includegraphics[width= \columnwidth]{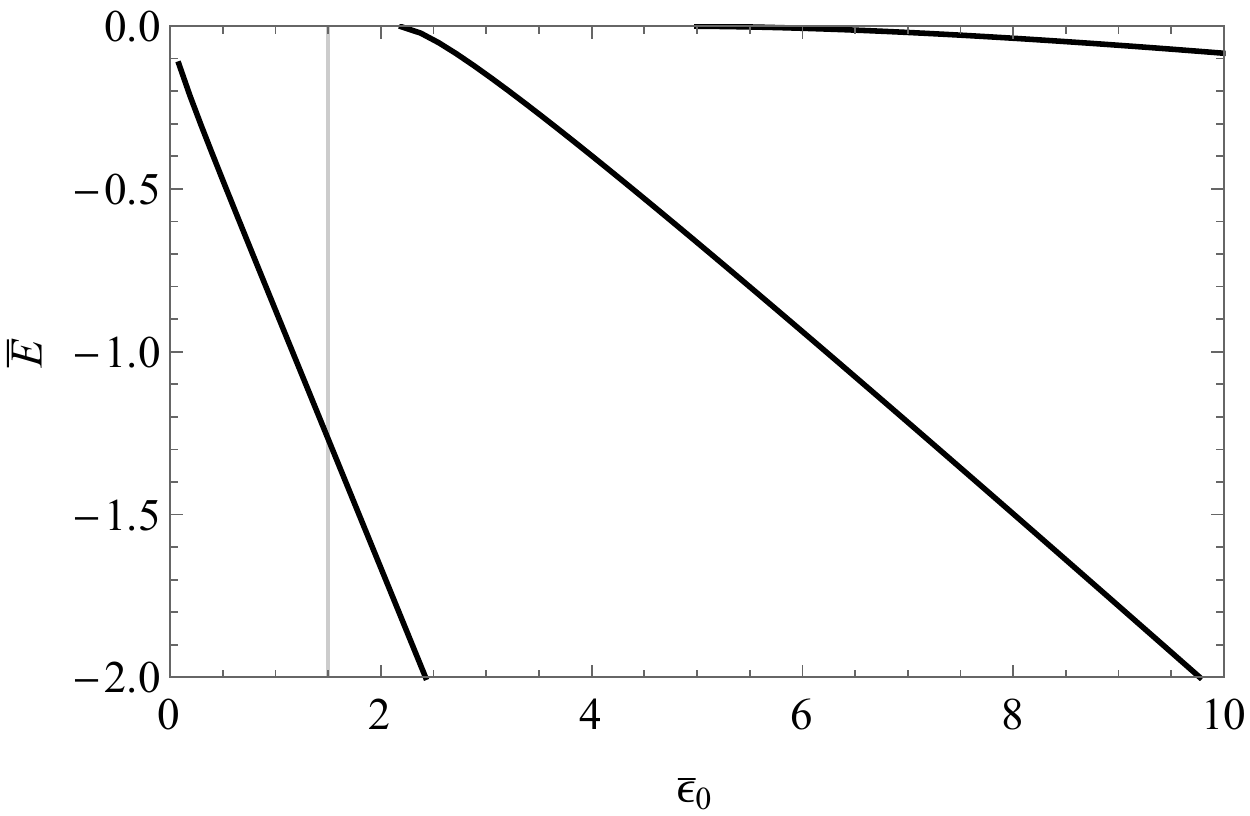}
\caption{Narrow resonance limit of the closed-channel ($\underline{ab}a$ and $\underline{aa}b$) trimer binding energies relative to the closed-channel dimer binding energy with respect to the closed-channel threshold, which in the narrow resonance limit coincides with the resonance position $\bar{\epsilon}_0$.
The gray line indicates the value $\bar{\epsilon}_0 = 1.5$ that we set the resonance position to throughout our analysis of the multichannel system with varying values of $\bar{r}^*$.}
\label{fig:clsoedchanneltrimers}
\end{figure}

However, since we fixed $\bar{\epsilon}_0 = 1.5$ the closed-channel system we consider here has a dimer binding energy of $\epsilon_0$ with respect to the closed-channel threshold in the limit $r^* \rightarrow \infty$.
In Fig.~\ref{fig:clsoedchanneltrimers} we show the closed-channel trimer spectrum in the narrow resonance limit with respect to the open-channel threshold as a function of the resonance position $\bar{\epsilon}_0$.
We observe that for increasing $\bar{\epsilon}_0$ the trimer binding energy increases and more trimer states are getting bound.
However, to keep the analysis for the coupled case as simple as possible, we have chosen the resonance position $\epsilon_0$ such that we just have a single non-universal closed-channel trimer state near the open-channel threshold in the limit of zero coupling to the open-channel. 
Hence, we have fixed the resonance position to $\bar{\epsilon}_0 = 1.5$ in Fig. \ref{fig:fig1}.

For completeness we note that the closed-channel spectrum related to the remaining part
\begin{equation}
\langle \underline{ab}b | \phi \rangle = \langle \underline{ab}b |K| \underline{ab}b  \rangle \langle \underline{ab}b | \phi \rangle
\end{equation}
is equivalent to the closed-channel one we just discussed, when $a_\mathrm{bg}$ is set to zero (compare Eqs.~(\ref{eq:aabHamiltonian}) and (\ref{eq:abbHamiltonian}) in appendix~\ref{app:Second Quantization}).
Since the $\underline{ab}b$ threshold energy lies $\epsilon_0$ higher in energy as the  $\underline{ab}a$ threshold for $\epsilon = \epsilon_0$ also the trimer spectrum needs to be shifted to higher energies by the threshold difference $\epsilon_0$.
Since the trimer binding energy $\bar{E}<1.5$ for $\bar{\epsilon}_0 = 1.5$ the $\underline{ab}b$ closed-channel trimer is located above the $\underline{aa}a$ threshold in the considered system. 
This is confirmed by our calculations that show only a single background trimer state in the narrow resonance limit.

\section{Conclusion and Outlook}
\label{sec:Conclusion and Outlook}
We present a multichannel version of the three-body bound state equations in momentum space.
In order to probe multichannel effects we analyze two different three-body realizations of an interaction, that on the two-body level leads to a standard two-channel model for Feshbach resonances with separable $s$-wave interactions. 
The two models, that correspond to the different realizations, differ only in the combination of single-particle spins employed for the closed channel of the two-body model. 
Realistic full coupled-channels models for atomic $s$-wave interactions include symmetric spin combinations of the form $| \mathpzc{c} \mathpzc{c} \rangle$ and $| \mathpzc{c} \mathpzc{c}' \rangle_S$ for the closed channels.
We analyze the three-body bound state spectrum for either a purely $|\mathpzc{c} \mathpzc{c} \rangle$ or a purely $| \mathpzc{c} \mathpzc{c}' \rangle_S$ realization of the closed channel for various values of the resonance width. 
We find that the realization of the interaction in spin space can strongly affect the Efimov spectrum for intermediate resonance widths.
Therefore our findings suggest that in this regime a full multichannel model is needed to identify the three-body parameter for identical bosonic alkali-metal atoms correctly. 
Additionally we find that trimers related to the closed channels can appear in the $| \mathpzc{c} \mathpzc{c}' \rangle_S$ realization.
Contrary to the regime of intermediate resonance width, we find that both the $|\mathpzc{c} \mathpzc{c} \rangle$ as well as the $| \mathpzc{c} \mathpzc{c}'\rangle_S$ configurations reduce to the same narrow resonance limit of the Efimov spectrum. 
In this limit,
the three-body dissociation scattering lengths $a_-^{(n)}$ scale linearly with the resonance width parameter $r^*$.
We derive this limiting behavior 
by analyzing the three-body bound state equation.
We find that the narrow resonance limit is independent of the interaction potential used.
For the scenarios investigated and in the broad resonance limit the excited Efimov states appear to be independent of the closed channel configuration and agree with predictions from the corresponding single-channel model.

Our model can be extended by adding more separable terms even with higher partial wave components to represent the long-range van der Waals tail of the atomic interactions correctly. 
Studying such a class of models with only two internal spin states could help to understand the effects leading to van der Waals universality in the multichannel system and might lead to a better understanding of the robustness of van der Waals universality for Feshbach resonances with intermediate resonance widths. 
For completeness we note that in a realistic system the channels $| \mathpzc{c} \mathpzc{c}' \rangle_S$ can be such that both $\mathpzc{c}$ and $\mathpzc{c}'$ are different from the incoming channel \cite{Secker:2020b}. 
However, to realize this situation a third internal state on the single atom level would be required.

\section*{Acknowledgements}
We thank Jinglun Li, Victor Colussi, Gijs Groeneveld, and Silvia Musolino for discussions.
This research is financially supported by the
Netherlands Organisation for Scientific Research (NWO)
under Grant No. 680-47-623.

\appendix
\section{Permutation operators and spin projection}
\label{app:Permutation operators and spin projection}
We can split $P_+^s$ in four components, by introducing the operators $\mathcal{P}_S$ and $\mathcal{P}_A$, which project the state onto either the symmetric or antisymmetric spin combinations of particle $i$ and $j$.
We find that 
\begin{align}
\mathcal{P}_S P_+^{s} \mathcal{P}_S &= \mathcal{P}_S P_-^{s} \mathcal{P}_S &=: P_{SS} \\
\mathcal{P}_A P_+^{s} \mathcal{P}_A &= \mathcal{P}_A P_-^{s} \mathcal{P}_A &=: P_{AA} \\
\mathcal{P}_S P_+^{s} \mathcal{P}_A &= - \mathcal{P}_S P_-^{s} \mathcal{P}_A &=: P_{SA}\\
\mathcal{P}_A P_+^{s} \mathcal{P}_S &= - \mathcal{P}_A P_-^{s} \mathcal{P}_S &=: P_{AS} \, 
\end{align}
We define the symmetric and antisymmetric spin bases as
\begin{align}
&| \mathpzc{c}' \odot \mathpzc{c}'', \mathpzc{c} \rangle \nonumber \\
&=  \frac{ | \mathpzc{c}' \rangle_1 \otimes | \mathpzc{c}'' \rangle_2 \otimes | \mathpzc{c} \rangle_3 +   | \mathpzc{c}'' \rangle_1 \otimes |\mathpzc{c}' \rangle_2 \otimes | \mathpzc{c} \rangle_3}{\sqrt{2( 1 + \delta_{\mathpzc{c}' \mathpzc{c}''})}} \\
&|  \mathpzc{c}' \wedge \mathpzc{c}'', \mathpzc{c} \rangle \nonumber \\
&=  \frac{  | \mathpzc{c}' \rangle_1 \otimes | \mathpzc{c}'' \rangle_2 \otimes| \mathpzc{c} \rangle_3 - | \mathpzc{c}'' \rangle_1 \otimes | \mathpzc{c}' \rangle_2 \otimes | \mathpzc{c} \rangle_3}{\sqrt{2}} 
\end{align}
One can then work out the expressions for $P_{SS}$, $P_{AA}$, $P_{AS}$ and $P_{SA}$ explicitly
\begin{align}
P_{SS} |  \mathpzc{c} \odot \mathpzc{c}', \mathpzc{c}'' \rangle &=  \frac{1}{2\sqrt{1+\delta_{\mathpzc{c} \mathpzc{c}'}}}\left( \sqrt{1+\delta_{\mathpzc{c}'' \mathpzc{c}}} |  \mathpzc{c}'' \odot \mathpzc{c},\mathpzc{c}' \rangle \right. \nonumber \\
&\phantom{=} \qquad \left. + \sqrt{1+\delta_{\mathpzc{c}'' \mathpzc{c}'}} |  \mathpzc{c}'' \odot \mathpzc{c}', \mathpzc{c} \rangle \right) \\
P_{AA} |  \mathpzc{c} \wedge \mathpzc{c}', \mathpzc{c}'' \rangle &=  \frac{1}{2}\left( | \mathpzc{c}'' \wedge \mathpzc{c} , \mathpzc{c}' \rangle - |  \mathpzc{c}'' \wedge \mathpzc{c}' , \mathpzc{c} \rangle \right) \\
P_{AS} | \mathpzc{c} \odot \mathpzc{c}', \mathpzc{c}'' \rangle &=  \frac{1}{2\sqrt{1+\delta_{\mathpzc{c} \mathpzc{c}'}}}\left( | \mathpzc{c}'' \wedge \mathpzc{c}, \mathpzc{c}' \rangle \right. \nonumber \\
&\phantom{=} \qquad \left. +  | \mathpzc{c}'' \wedge \mathpzc{c}', \mathpzc{c} \rangle \right) \\
P_{SA} | \mathpzc{c} \wedge \mathpzc{c}', \mathpzc{c}'' \rangle &=  \frac{1}{2}\left( \sqrt{1+\delta_{\mathpzc{c}'' \mathpzc{c}}} | \mathpzc{c}'' \odot \mathpzc{c}, \mathpzc{c}' \rangle \right. \nonumber \\
&\phantom{=} \qquad \left. - \sqrt{1+\delta_{\mathpzc{c}'' \mathpzc{c}'}} |  \mathpzc{c}'' \odot \mathpzc{c}', \mathpzc{c} \rangle \right)\, .
\end{align}
\section{The transition matrix for the bb-channel configuration}
\label{app:The transition matrix for the bb-channel configuration}

The two-body $t$-operator can be defined as
\begin{equation}
t(z) = V + V g(z) V \, ,
\end{equation}
where $g(z) = (z - H)^{-1}$ is the Green's operator related to the relative two-body Hamiltonian. 
This implies that the open channel component of the transition operator $t_{PP}(z)$ can be expressed as 
\begin{align}
\label{eq:tppfirst}
t_{PP}(z) &= V_{PP}\left(1+g_{PP} V_{PP}+g_{PQ}V_{QP}\right)\notag \\
&+V_{PQ}\left(g_{QP}V_{PP}+g_{QQ}V_{QP}\right).
\end{align}
In order to simplify this expression we derive an operator version of the Feshbach formalism to eliminate the contributions $g_{QP}$, $g_{PQ}$ and $g_{QQ}$ in Eq. \eqref{eq:tppfirst}, we apply the definition $g(z)(z-H) \equiv 1$, such that we find 
\begin{align}
\label{eq:Gqq}
g_{QQ} = g^0_{QQ}+g_{QP}V_{PQ}g^0_{QQ},
\end{align}
with $g^0_{QQ}(z) = (z-H_{QQ})^{-1}$ and 
\begin{align}
\label{eq:Gpq,Gqp}
V_{PQ}g_{QP} = g_{PQ}V_{QP} = g_{PP}W_{PP},
\end{align}
where we have introduced the factor $W_{PP} = V_{PQ}g^0_{QQ}(z)V_{QP}$.  
Substituting Eqs. \eqref{eq:Gqq} and \eqref{eq:Gpq,Gqp} into Eq. \eqref{eq:tppfirst}, we obtain
\begin{align}
\label{eq:tppsecond}
t_{PP}(z) = \left(V_{PP}+W_{PP}\right)\left[1+g_{PP}\left(V_{PP}+W_{PP}\right)\right].
\end{align}
We recognize that Eq. \eqref{eq:tppsecond} looks like a single-channel transition operator where the open channel potential interaction strength $V_{PP}$ has been replaced by an effective interaction strength $V_{PP}+W_{PP}$. \par
Proceeding with the analysis of Eq. \eqref{eq:tppsecond}, we introduce the uncoupled transition operator $t_{unc}^{PP}$, defined as 
\begin{align}
\label{eq:Tuncoupled}
t_{PP}^{unc} = V_{PP}+V_{PP}g_0 t^{unc}_{PP},
\end{align}
such that we can rewrite Eq. \eqref{eq:tppsecond} as
\begin{align}
\label{eq:tppthird}
t_{PP} = &t_{PP}^{unc}+\left(1+V_{PP} g^0_{PP} \right)W_{PP}  \notag \\[3pt]
& \times \left(1+g_{PP}\left(V_{PP}+W_{PP}\right)\right).
\end{align}
Equation \eqref{eq:tppthird} can be simplified through the application of the single resonance approximation. Under this approximation, the resolvent operator $g^0_{QQ}$ can be replaced by its dominant contribution, such that $g^0_{QQ} \approx (E-E_\mathrm{b})^{-1}\ket{\phi_Q}\bra{\phi_Q}$, with closed channel bound state $\ket{\phi_Q}$ and binding energy $E_\mathrm{b}$.
Substituting this form of this Green's function into the definition of the potential operator $W_{PP}$, Eq. \eqref{eq:tppthird} reduces to 
\begin{align}
\label{eq:tpp4}
t_{PP} = &t_{PP}^{unc} +\frac{1}{{(E-E_\mathrm{b})}}\left(1+V_{PP} g^0_{PP} \right)V_{PQ}\ket{\phi_Q} \nonumber\\
& \times \bra{\phi_Q}V_{QP}\left(1+g_{PP}\left(V_{PP}+W_{PP}\right)\right).
\end{align}
Using the resolvent equation for $g_{PP}=g_{PP}^0 + g_{PP}^0 W_{PP} g_{PP}$ (which can be derived from the expression for $g_{PP}$ analogous to Eq. (\ref{eq:Gqq}) in combination with Eq. (\ref{eq:Gpq,Gqp})) the last term can be rewritten as follows
\begin{align}
&\bra{\phi_Q}V_{QP}\left(1+g_{PP}\left(V_{PP}+W_{PP}\right)\right) \nonumber\\
&=\bra{\phi_Q}V_{QP}\left(1 + g_{PP}^0 V_{PP} \right.\\
&\phantom{=} \left. + g_{PP}^0 W_{PP}\left(1+g_{PP}\left(V_{PP}+W_{PP}\right)\right)\right) \nonumber \, .
\end{align}
Replacing the $g^0_{QQ}$ by the approximation we used earlier in the first $W_{PP}$ of the last line we arrive at an equation, which we can solve for 
\begin{align}
& \bra{\phi_Q}V_{QP}\left(1+g_{PP}\left(V_{PP}+W_{PP}\right)\right) \nonumber \\
&= \frac{(E-E_\mathrm{b})\bra{\phi_Q}V_{QP}\left(1 + g_{PP}^0 V_{PP} \right)}
{E - E_\mathrm{b} - \bra{\phi_Q}V_{QP}g^0_{PP}V_{PQ} \ket{\phi_Q}}
\end{align}
such that we find
\begin{widetext}
\begin{align}
\label{eq:tpp5}
t_{PP} &= t_{PP}^{unc} + \frac{\left(1+V_{PP} g^0_{PP} \right)V_{PQ}\ket{\phi_Q}\bra{\phi_Q}V_{QP}\left(1 + g_{PP}^0 V_{PP} \right)}
{E - E_\mathrm{b} - \bra{\phi_Q}V_{QP}g^0_{PP}V_{PQ} \ket{\phi_Q}} \\
& = t_{PP}^{unc} + \frac{\left(1+V_{PP} (g_{0} + g_{0} t_{PP}^{unc} g_{0}) \right)V_{PQ}\ket{\phi_Q}\bra{\phi_Q}V_{QP}\left(1 + (g_{0} + g_{0} t_{PP}^{unc} g_{0}) V_{PP} \right)}
{E - E_\mathrm{b} - \bra{\phi_Q}V_{QP}(g_{0} + g_{0} t_{PP}^{unc} g_{0})V_{PQ} \ket{\phi_Q}} \, .
\end{align}
\end{widetext}
Where we replaced the Green's functions $g^0_{PP}$ with the identity 
\begin{align}
\label{eq:Gpp0}
g^0_{PP} = g_{0} + g_{0} t_{PP}^{unc} g_{0}.
\end{align}
We can now use the separable interaction to explicitly get
\begin{equation}
t_{PP}^{unc} = \tau_{PP}^{unc} | \zeta \rangle \langle \zeta |
\end{equation}
with
\begin{equation}
\tau_{PP}^{unc} = \frac{\bar{v}_{PP}}{m \Lambda -\bar{v}_{PP} \langle \zeta | g_0 | \zeta \rangle} \, .
\end{equation}
The interaction strength $\bar{v}_{PP}$ is then related to the background scattering length
\begin{equation}
\label{eq:vPPabg}
\bar{v}_{PP} = \frac{\bar{a}_\mathrm{bg}}{2 \pi^2 + \bar{a}_\mathrm{bg} \overline{\langle \zeta | g_0(0) | \zeta \rangle}} \equiv  \frac{\bar{a}_\mathrm{bg}}{2 \pi^2}\Gamma \, .
\end{equation}
Note that the bar indicates quantities made dimensionless in units of $\Lambda$ as introduced in the beginning of section \ref{sec:Results}.
We define $\xi (z) = \langle \zeta | g_0(z) | \zeta \rangle$ and have that $t_{PP} = \tau_{PP} | \zeta \rangle \langle \zeta |$, since $|\zeta \rangle$ also appears in $V_{PQ}$ and find 
\begin{widetext}
\begin{align}
\tau_{PP} &= \tau_{PP}^{unc} + \frac{\left(1+v_{PP} (\xi (z) + \xi (z) \tau_{PP}^{unc} \xi (z)) \right)v_{PQ}\braket{\zeta | \phi_Q}\braket{\phi_Q|\zeta}v_{QP}\left(1 + (\xi (z) + \xi (z) \tau_{PP}^{unc} \xi (z)) v_{PP} \right)}
{z - E_\mathrm{b} - \braket{\phi_Q | \zeta}v_{QP}(\xi (z) + \xi (z) \tau_{PP}^{unc} \xi (z))v_{PQ} \braket{\zeta|\phi_Q}} \\
&= \tau_{PP}^{unc} + \frac{\left(1+v_{PP} (\xi  + \xi^2 \tau_{PP}^{unc}) \right)^2 |v_{PQ}\braket{\zeta | \phi_Q}|^2}
{z - E_\mathrm{b} - (\xi + \xi^2 \tau_{PP}^{unc})|v_{PQ}\braket{\zeta | \phi_Q}|^2} \, . \label{eq:taufeshbachfinal}
\end{align}  
\end{widetext}
In this final form $t_{PP}$ is solely determined by $\bar{v}_{PP}$, $\bar{v}_{PQ}$ and $E_\mathrm{b}$,
which can be related to $a_\mathrm{bg}$ (Eq.~\eqref{eq:vPPabg}) and $r^*$ by considering the $z \rightarrow 0$ limit
\begin{equation}
\bar{a} = 2 \pi^2 \bar{\tau}_{PP}(0)|\zeta(0)|^2 = \bar{a}_\mathrm{bg} + \frac{ 2 \pi^2 |\zeta(0)|^2  \bar{g}^2 /\Gamma^{2} }{-\bar{E_\mathrm{b}}-\bar{\xi} (0) \bar{g}^2 / \Gamma}
\end{equation}
such that we arrive at
\begin{equation}
\bar{g} \equiv \overline{|v_{PQ}\braket{\zeta | \phi_Q}|} = \frac{\Gamma}{\sqrt{2 \pi^2 \bar{r}^*} |\zeta(0)|}\, .
\end{equation}
Fixing $\tilde{a} = \bar{a} / \bar{r}^*$ and $\bar{a}_\mathrm{bg}$ we can find $\bar{E}_\mathrm{b}$ in the narrow resonance limit
\begin{equation}
-\bar{E}_\mathrm{b} \underset{\bar{r}^* \rightarrow \infty}{=} \frac{1}{\bar{r}^* \tilde{a}} + \bar{\xi}(0)\bar{g}^2/\Gamma 
\, .
\end{equation}
We replace $-\bar{E}_\mathrm{b}$ by this expression in $\bar{\tau}_{PP}$ and change to units determined by $r^*$, that we indicate by a tilde. 
We have $\bar{z}=\tilde{z}/\bar{r}^{*2}$ and therefore $\bar{z}\rightarrow 0$ in the narrow resonance limit.
This leads us to 
\begin{widetext}
\begin{align}
\tilde{\tau}_{PP} &\underset{\phantom{\bar{r}^* \rightarrow \infty}}{\equiv} \bar{\tau}_{PP}/\bar{r}^* \\
&\underset{\phantom{\bar{r}^* \rightarrow \infty}}{=} \bar{\tau}_{PP}^{unc}/\bar{r}^*+\frac{1}{2 \pi^2 |\zeta(0)|^2 \bar{r}^{*2}} \times \frac{1+\mathcal{O}(\sqrt{-\bar{z}})}{\bar{z} +\frac{1}{\bar{r}^{*2} \tilde{a}}+\left[\bar{g}^2\left(\bar{\xi}(0)+\bar{\xi}(0)^2 \bar{\tau}^{unc}_{PP}(0) -\bar{\xi}(\bar{z})-\bar{\xi}(\bar{z})^2 \bar{\tau}^{unc}_{PP}(\bar{z})\right) \right]} \\
&\underset{\phantom{\bar{r}^* \rightarrow \infty}}{=} \bar{\tau}_{PP}^{unc}/\bar{r}^* + \frac{1}{2 \pi^2 |\zeta(0)|^2 \bar{r}^{*2}} \times \frac{1+\mathcal{O}(\sqrt{-\bar{z}})}{\bar{z} +\frac{1}{\bar{r}^{*2} \tilde{a}}+\left[\frac{-\sqrt{-\bar{z}} + \mathcal{O}(\sqrt{-\bar{z}}^2)}{\bar{r}^*}\right]} \\
&\underset{\bar{r}^* \rightarrow \infty}{=} \frac{1}{2 \pi^2 |\zeta(0)|^2 } \times \frac{1}{\tilde{z} +\tilde{a}^{-1}-\sqrt{-\tilde{z}}} \, . \label{eq:tppNarrowLimitApp}
\end{align}
\end{widetext}
In the first step we consider the numerator in the second term of Eq.~\eqref{eq:taufeshbachfinal}. We replaced $\bar{\xi}$ by the limiting expression $\bar{\xi}(\bar{z}) \approx 2 \pi^2 \sqrt{-\bar{z}} |\zeta(0)|+\bar{\xi}(0)$ for small $\sqrt{-\bar{z}}$.
With that we taylor expanded the numerator around $\sqrt{-\bar{z}} = 0$.
The zeroth order term is then simply the $\bar{z} \rightarrow 0$ limit $(2 \pi^2 |\zeta(0)|^2 \bar{r}^{*2})^{-1}$.
In the second step we consider the term $\left[ ... \right]$ and proceed similarly to  arrive at the lowest order term in $\sqrt{-\bar{z}}$.
We have that $\bar{\tau}_{PP}^{unc}/\bar{r}^* \rightarrow 0$ in the narrow resonance limit and rewrite $\bar{z}=\tilde{z}/\bar{r}^{*2}$.
With that we can take the limit $\bar{r}^* \rightarrow \infty$ in the final step.
Equation \eqref{eq:tppNarrowLimitApp} corresponds to Eq. \eqref{eq:tppNarrowLimit} as presented in the main text.

\begin{widetext}
\section{Second Quantization}
\label{app:Second Quantization}
We derive the second quantized form of the $Q = \underline{bb} $ and $Q = \underline{ab} $ realization of the three-body systems considered to be able to relate them to the Hamiltonians commonly considered in effective field theories. 
The notation we use in this section deviates from the one in the main text with $\alpha = (i,\sigma)$ we now label a base $|i \rangle |\sigma \rangle$ in the single particle Hilbert space $\mathcal{H}_1$, where $\sigma$ labels the internal or spin degrees of freedom and $i $ a base in configuration space.

\subsection{Creation and Annihilation Operators}
We have the bosonic annihilation and creation operators $a_\alpha$ and $a^\dag_\alpha$ with 
\begin{align}
\left[ a_\alpha, a_\beta^\dag  \right] &= \delta_{\alpha \beta} \nonumber
\\
\left[ a_\alpha, a_\beta  \right] &= 0
\\
\left[ a_\alpha^\dag, a_\beta^\dag  \right] &= 0 \nonumber
\end{align}
The operators $a_\alpha$ and $a^\dag_\alpha$ act on the symmetric part of Fock space $\mathcal{P}_S \mathcal{H}_\mathcal{F}$
\begin{equation}
\mathcal{H}_\mathcal{F} = \bigoplus_N \mathcal{H}_1^N \nonumber
\end{equation}
by
\begin{equation}
a_{\alpha} | 0 \rangle = 0
\end{equation}
and
\begin{equation}
a^\dag_{\alpha_1} \dots a^\dag_{\alpha_N} | 0 \rangle = \frac{1}{\sqrt{N!}} \sum_{\mathfrak{p} \in \Sigma(N)} P_\mathfrak{p} | \alpha_1 \dots \alpha_N \rangle \, ,
\end{equation}
with $\Sigma(N)$ the permutation group of $N$ elements and $P_\mathfrak{p}$ the permutation operator related to the permutation $\mathfrak{p}$. 
The projector on the totally symmetric subspace is defined as
\begin{equation}
\mathcal{P}^N_S = \frac{1}{N!} \sum_{\mathfrak{p} \in \Sigma(N)} P_\mathfrak{p}
\end{equation}

\subsection{Kinetic Energy and Interaction Potential}

The kinetic energy operator $T$ acts on a single particle as
\begin{equation}
T_{1b} = \sum_{\alpha \beta} \langle \alpha | T | \beta \rangle |\alpha \rangle \langle \beta |
\end{equation}
with 
$\langle \alpha | T | \beta \rangle = \langle i_\alpha |- \hbar^2 \Delta /2m + E_{\sigma_\alpha}) | i_\beta \rangle \delta_{\sigma_\alpha \sigma_\beta}$.
The interaction potential $V$ acts on two particles as
\begin{equation}
V_{2b} =  \sum_{\alpha_1 \alpha_2 \beta_1 \beta_2} \langle \alpha_1 \alpha_2 | V | \beta_1 \beta_2 \rangle | \alpha_1 \alpha_2 \rangle \langle \beta_1 \beta_2 | \, ,
\end{equation}
with $\langle \alpha_1 \alpha_2 | V | \beta_1 \beta_2 \rangle = \langle i_{\alpha_1} i_{\alpha_2} | V_{\sigma_{\alpha_1}\sigma_{\alpha_2}, \sigma_{\beta_1}\sigma_{\beta_2} } | i_{\beta_1} i_{\beta_2} \rangle$.
For three particles restricting to the fully symmetric subspace we have the kinetic energy operator
\begin{align}
T_{3b} \mathcal{P}_S^3 & = (T_{1b}^1 + T_{1b}^2 + T_{1b}^3) \mathcal{P}_S^3 \nonumber \\
& =\frac{1}{2}(1 + P_+ + P_-)(1 + P_{23})T_{1b}^1 \mathcal{P}_S^3 \nonumber\\
& = \sum_{\underline{\alpha} \underline{\beta}} \frac{3!}{2}\langle \alpha_1 | T | \beta_1 \rangle \delta_{\alpha_2 \beta_2} \delta_{\alpha_3 \beta_3} \mathcal{P}_S^3 |\underline{\alpha} \rangle \langle \underline{\beta}| \mathcal{P}_S^3 \\
& = \sum_{\underline{\alpha} \underline{\beta}} \frac{1}{2}\langle \alpha_1 | T | \beta_1 \rangle \delta_{\alpha_2 \beta_2} \delta_{\alpha_3 \beta_3} a^\dag_{\alpha_1} a^\dag_{\alpha_2} a^\dag_{\alpha_3} |0 \rangle \langle 0 | a_{\beta_1} a_{\beta_2} a_{\beta_3} \nonumber\\
& = \sum_{\alpha \beta} \langle \alpha | T | \beta \rangle a^\dag_{\alpha} a_{\beta} \, , \nonumber
\end{align}
with $P_{23}$ the permutation exchanging particles $2$ and $3$ and $\underline{\alpha} $ a shorthand notation for the three indices $(\alpha_1 \alpha_2 \alpha_3)$.
The interaction term then is
\begin{align}
V_{3b} \mathcal{P}_S^3 & = (V^{23} + V^{31} + V^{12}) \mathcal{P}_S^3 \nonumber \\
& =\frac{1}{2}(1 + P_+ + P_-)(1 + P_{23})V^{23} \mathcal{P}_S^3 \nonumber\\
& = \sum_{\underline{\alpha} \underline{\beta}} \frac{3!}{2}\langle \alpha_1 \alpha_2 | V | \beta_1 \beta_2\rangle \delta_{\alpha_3 \beta_3} \mathcal{P}_S^3 |\underline{\alpha} \rangle \langle \underline{\beta}| \mathcal{P}_S^3 \\
& = \sum_{\underline{\alpha} \underline{\beta}} \frac{1}{2}\langle \alpha_1 \alpha_2 | V | \beta_1 \beta_2\rangle \delta_{\alpha_3 \beta_3} a^\dag_{\alpha_1} a^\dag_{\alpha_2} a^\dag_{\alpha_3} |0 \rangle \langle 0 | a_{\beta_1} a_{\beta_2} a_{\beta_3} \nonumber \\
& = \sum_{\alpha_1 \alpha_2 \beta_1 \beta_2} \frac{1}{2}\langle \alpha_1 \alpha_2 | V | \beta_1 \beta_2 \rangle a^\dag_{\alpha_1} a^\dag_{\alpha_2} a_{\beta_1} a_{\beta_2} \, . \nonumber
\end{align}

\subsection{The $Q = \underline{bb}$ and $Q = \underline{ab}$ realization}
We consider a model with two internal states per particle $ \sigma \in \{ a, b \}$. We find that
\begin{equation}
T_{3b} \mathcal{P}_S^3 = \sum_{i j} \left[ \langle i |- \hbar^2 \Delta /2m + E_{a} | j \rangle a^\dag_{i} a_{j} + \langle i |- \hbar^2 \Delta /2m + E_{b} | j \rangle b^\dag_{i} b_{j}\right] \, ,
\end{equation}
with $a_i = a_{(i,a)}$ and $b_i = a_{(i,b)}$.
Furthermore we restrict to an interaction term which reduces to a simple two-channel model on the two particle subspace. 

\subsubsection{Explicit Representations of the Field Operators}
There is a unitary transformation 
\begin{equation}
U:\mathcal{P}_S^3 \mathcal{H}_1^3 \rightarrow \left[\mathcal{P}^3_{S,aaa}\mathcal{H}_{1,a}^3 \right] \oplus \left[ \mathcal{H}_{1,b} \otimes \left(\mathcal{P}_{S,aa}^2\mathcal{H}_{1,a}^2 \right) \right] \oplus \left[ \mathcal{H}_{1,a} \otimes \left(\mathcal{P}_{S,bb}^2\mathcal{H}_{1,b}^2 \right) \right] \oplus \left[\mathcal{P}^3_{S,bbb}\mathcal{H}_{1,b}^3 \right]
\end{equation}
connecting the representations of the field operator algebra. It is defined naturally by
\begin{equation}
a^\dag_\alpha a^\dag_\beta a^\dag_\gamma | 0 \rangle \mapsto 
\begin{cases}
a^\dag_{i_\alpha} a^\dag_{i_\beta} a^\dag_{i_\gamma} | 0 \rangle & \text{ for } \sigma_\alpha = \sigma_\beta = \sigma_\gamma = a \, , \\
a^\dag_{i_\alpha} a^\dag_{i_\beta} b^\dag_{i_\gamma} | 0 \rangle & \text{ for } \sigma_\alpha = \sigma_\beta = a \text{ and } \sigma_\gamma = b \, , \\
a^\dag_{i_\alpha} b^\dag_{i_\beta} b^\dag_{i_\gamma} | 0 \rangle & \text{ for } \sigma_\alpha = a \text{ and } \sigma_\beta = \sigma_\gamma = b \, , \\
b^\dag_{i_\alpha} b^\dag_{i_\beta} b^\dag_{i_\gamma} | 0 \rangle & \text{ for } \sigma_\alpha = \sigma_\beta = \sigma_\gamma = b \, .
\end{cases}
\end{equation} 

\subsubsection{$Q = \underline{bb}$ realization}
First we consider the case 
\begin{equation}
\langle i_{\alpha_1} i_{\alpha_2} | V_{\sigma_{\alpha_1}\sigma_{\alpha_2}, \sigma_{\beta_1}\sigma_{\beta_2} } | i_{\beta_1} i_{\beta_2} \rangle = 0 \quad \forall \quad \left[(\sigma_{\alpha_1}\sigma_{\alpha_2}) \neq (aa) \text{ or } (bb) \right] \text{or} \left[ (\sigma_{\beta_1}\sigma_{\beta_2}) \neq (aa) \text{ or } (bb) \right]
\end{equation}
such that we have 
\begin{align}
V_{3b} \mathcal{P}_S^3 &= \sum_{i_1 i_2 j_1 j_2} \frac{1}{2} \left[ \langle i_1 i_2 | V_{aa,aa} | j_1 j_2 \rangle a^\dag_{i_1} a^\dag_{i_2} a_{j_1} a_{j_2} + \langle i_1 i_2 | V_{bb,bb} | j_1 j_2 \rangle b^\dag_{i_1} b^\dag_{i_2} b_{j_1} b_{j_2} \right.\\
& \phantom{=} \left. \qquad \qquad \qquad + \langle i_1 i_2 | V_{bb,aa} | j_1 j_2 \rangle b^\dag_{i_1} b^\dag_{i_2} a_{j_1} a_{j_2} + \langle i_1 i_2 | V_{aa,bb} | j_1 j_2 \rangle a^\dag_{i_1} a^\dag_{i_2} b_{j_1} b_{j_2} \right] \, . \nonumber
\end{align}
It is easy to show that there is no coupling between the subspaces spanned by
\begin{align}
| i j k \rangle_{aaa}^S &\sim a_i^\dag a_j^\dag a_k^\dag | 0 \rangle \\
| i j k \rangle_{abb}^S &\sim a_i^\dag b_j^\dag b_k^\dag | 0 \rangle  \nonumber
\end{align}
and 
\begin{align}
| i j k \rangle_{bbb}^S &\sim b_i^\dag b_j^\dag b_k^\dag | 0 \rangle \\
| i j k \rangle_{aab}^S &\sim a_i^\dag a_j^\dag b_k^\dag | 0 \rangle \, . \nonumber
\end{align}
We can rewrite the Hamiltonian acting in the $aaa$ and $abb$ channels as
\begin{equation}
H = H_{aaa,aaa} + H_{abb,aaa} + H_{aaa,abb} + H_{abb,abb} \, ,
\end{equation}
with 
\begin{align}
H_{aaa,aaa} &=  \sum_{\underline{i} \underline{j}}\left[ \frac{1}{2} \langle i_1 |T_{a} | j_1 \rangle \delta_{i_2 j_2} \delta_{i_3 j_3} + \frac{1}{2} \langle i_1 i_2 | V_{aa,aa} | j_1 j_2 \rangle \delta_{i_3 j_3} \right] a^\dag_{i_1} a^\dag_{i_2} a^\dag_{i_3} |0\rangle \langle 0 | a_{j_1} a_{j_2} a_{j_3} \\
H_{abb,abb} &=  \sum_{\underline{i} \underline{j}}\left[ \frac{1}{2} \langle i_1 |T_{a} | j_1 \rangle \delta_{i_2 j_2} \delta_{i_3 j_3} 
+ \langle i_2 |T_{b} | j_2 \rangle \delta_{i_1 j_1} \delta_{i_3 j_3} \right.  \\
& \phantom{=} \qquad \qquad \left. + \frac{1}{2} \langle i_2 i_3 | V_{bb,bb} | j_2 j_3 \rangle \delta_{i_1 j_1} \right] a^\dag_{i_1} b^\dag_{i_2} b^\dag_{i_3} |0\rangle \langle 0 | a_{j_1} b_{j_2} b_{j_3} \nonumber\\
H_{abb,aaa} & = \sum_{\underline{i} \underline{j}} \left[ \frac{1}{2} \langle i_2 i_3 | V_{bb,aa} | j_2 j_3 \rangle \delta_{i_1 j_1}  \right]
a^\dag_{i_1} b^\dag_{i_2} b^\dag_{i_3} |0\rangle \langle 0 | a_{j_1} a_{j_2} a_{j_3} \\
H_{aaa,abb} & = \sum_{\underline{i} \underline{j}} \left[ \frac{1}{2} \langle i_2 i_3 | V_{aa,bb} | j_2 j_3 \rangle \delta_{i_1 j_1}  \right]
a^\dag_{i_1} a^\dag_{i_2} a^\dag_{i_3} |0\rangle \langle 0 | a_{j_1} b_{j_2} b_{j_3} \, ,
\end{align}
where we again introduced $\underline{i}$ as a shorthand for the three indices $(i_1 i_2 i_3)$.
We get
\begin{align}
H_{aaa,aaa} &= \left(T_{a}^1 + T_{a}^2 + T_{a}^3 + V_{aa,aa}^{23} + V_{aa,aa}^{31} + V_{aa,aa}^{12} \right) \mathcal{P}_{S,aaa}^3 \\
H_{abb,abb} &= (1 + P_{23,abb})  \sum_{\underline{i} \underline{j}} \left( \frac{1}{2} \langle i_1 |T_{a} | j_1 \rangle \delta_{i_2 j_2} \delta_{i_3 j_3} 
+ \langle i_2 |T_{b} | j_2 \rangle \delta_{i_1 j_1} \delta_{i_3 j_3} \right. \nonumber\\
& \phantom{=} \qquad \qquad \left. + \frac{1}{2} \langle i_2 i_3 | V_{bb,bb} | j_2 j_3 \rangle \delta_{i_1 j_1} \right) |\underline{i}\rangle_{abb} {}_{abb}\langle \underline{j} | (1 + P_{23,abb})/2  \nonumber\\
& = (1 + P_{23,abb}) \left( \frac{1}{2} T_a^1
+ T_b^2 + \frac{1}{2} V_{bb,bb}^{23} \right) (1 + P_{23,abb})/2 \nonumber \\
& = \left( T_a^1
+ T_b^2 + T_b^3 + V_{bb,bb}^{23} \right) (1 + P_{23,abb})/2  \\
H_{abb,aaa} & = \sqrt{3} (1 + P_{23,abb})\sum_{\underline{i} \underline{j}} \left[ \frac{1}{2} \langle i_2 i_3 | V_{bb,aa} | j_2 j_3 \rangle \delta_{i_1 j_1}  \right] |\underline{i} \rangle_{abb} {}_{aaa}\langle \underline{j} | \mathcal{P}_{S,aaa}^3 \nonumber\\
& = \sqrt{3} V_{bb,aa}^{23}  \mathcal{P}_{S,aaa}^3 \\
H_{aaa,abb} & = \sqrt{2} \sqrt{6} \mathcal{P}_{S,aaa}^3 \sum_{\underline{i} \underline{j}} \left[ \frac{1}{2} \langle i_2 i_3 | V_{aa,bb} | j_2 j_3 \rangle \delta_{i_1 j_1}  \right]
| \underline{i} \rangle \langle \underline{j} | (1 + P_{23,abb})/2 \nonumber\\
& = \frac{1}{\sqrt{3}} (1 + P_{+,aaa} + P_{-,aaa})(1 + P_{23,aaa}) \sum_{\underline{i} \underline{j}} \left[ \frac{1}{2} \langle i_2 i_3 | V_{aa,bb} | j_2 j_3 \rangle \delta_{i_1 j_1}  \right]
| \underline{i} \rangle \langle \underline{j} | (1 + P_{23,abb})/2 \nonumber\\
& = \frac{1}{\sqrt{3}} (1 + P_{+,aaa} + P_{-,aaa}) V_{aa,bb}^{23} (1 + P_{23,abb})/2 \, ,
\end{align}
when representing the model on $\mathcal{H}_{1,a}^3 \oplus \mathcal{H}_{1,a} \otimes \mathcal{H}_{1,b}^2$, where the indices $aaa$ and $abb$ of the operators $P_+$, $P_-$ and $P_{23}$ indicate the subspaces they are acting on and $P_{23,abb} = 1_a \otimes \mathcal{P}_{S,bb}^2 $.

In the following we approximate the Hamiltonian $H_{abb,abb}$ by introducing a dimer field.
For that we introduce relative $r$ and center-of-mass coordinates $R$ for the two particles in the $b$ state with positions $r_2$ and $r_3$.
We assume the following form of the interaction  $V_{bb,bb}^{23} = 1_{r_1} \otimes 1_R \otimes V_{bb,bb}^r $ with $\psi_E$ a bound Eigenstate fulfilling $ (-\hbar \Delta_r /m + V_{bb,bb}^r)\psi_E = E \psi_E$.
We introduce the projector $\mathcal{P}_{\psi_E}$ onto the state $\psi_E$ such that we can restrict $H_{abb,abb}$ to $H_{abb,abb} \mathcal{P}_{\psi_E} = \mathcal{P}_{\psi_E} H_{abb,abb} = H_{abb,abb}^{\psi_E} $.
The coupling term then acts on a state $\phi$ as $[\mathcal{P}_{\psi_E} V_{bb,aa}^{23} \psi](r_1,r_2,r_3) = \psi_E(r_2 - r_3) \int d r \chi^*(r) \phi(r_1,(r_2 +r_3)/2 + r, (r_2 + r_3)/2 - r) $ with $\chi^*(r) = [\psi_E^* V_{bb,aa}](r)$.
With that we can approximate the total Hamiltonian by
\begin{align}
H_{aaa,aaa} &= \left(T_{a}^1 + T_{a}^2 + T_{a}^3 + V_{aa,aa}^{23} + V_{aa,aa}^{31} + V_{aa,aa}^{12} \right) \mathcal{P}_{S,aaa}^3 \\
H_{abb,abb}
& \approx \left( T_a^1
+ T_d + E \right) \mathcal{P}_{\psi_E} \\
H_{abb,aaa} 
& \approx \sqrt{3} \mathcal{P}_{\psi_E} V_{bb,aa}^{23}  \mathcal{P}_{S,aaa}^3 \\
H_{aaa,abb}
& \approx \frac{1}{\sqrt{3}} (1 + P_{+,aaa} + P_{-,aaa}) V_{aa,bb}^{23} \mathcal{P}_{\psi_E}
\end{align}
with $T_d = - \hbar^2 \Delta_R / 4m$.
By rewriting the projector 
\begin{align}
\mathcal{P}_{\psi_E} \left( 1 + P_{23,abb} \right)/2 &= \left( 1 + P_{23,abb} \right)/2 \sum_{\underline{i} \underline{j} \tilde{k}} \langle i_2 i_3|\psi_E, \tilde{k} \rangle \langle \psi_E, \tilde{k} | j_2 j_3 \rangle \delta_{i_1 j_1} |\underline{i} \rangle_{abb} {}_{abb}\langle \underline{j} |\left( 1 + P_{23,abb} \right)/2 \\
& = \sum_{\underline{i} \underline{j} \tilde{k}} \left[ \frac{1}{2} \langle i_2 i_3|\psi_E, \tilde{k} \rangle \langle \psi_E, \tilde{k} | j_2 j_3 \rangle \delta_{i_1 j_1} \right] a^\dag_{i_1} b^\dag_{i_2} b^\dag_{i_3} |0\rangle \langle 0 | a_{j_1} b_{j_2} b_{j_3} \\
& = \sum_{k \tilde{k}} a^\dag_{k} \left( \sum_{i_2 i_3} \left[ \frac{1}{\sqrt{2}} \langle i_2 i_3|\psi_E, \tilde{k} \rangle b^\dag_{i_2} b^\dag_{i_3} \right] \right)  |0\rangle \langle 0 | a_{k} \left( \sum_{j_2 j_3} \left[ \frac{1}{\sqrt{2}} \langle \psi_E, \tilde{k} | j_2 j_3 \rangle  b_{j_2} b_{j_3} \right] \right)  \\
& = \sum_{k \tilde{k}} a^\dag_{k} d^\dag_{\tilde{k}} |0\rangle \langle 0 | a_{k} d_{\tilde{k}} \, ,
\end{align}
with $\tilde{k}$ labeling a base in $R$
we can single out the dimer field operator $d_{\tilde{k}}$.
We find
\begin{align}
H_{aaa,aaa} &=  \sum_{\underline{i} \underline{j}}\left[ \frac{1}{2} \langle i_1 |T_{a} | j_1 \rangle \delta_{i_2 j_2} \delta_{i_3 j_3} + \frac{1}{2} \langle i_1 i_2 | V_{aa,aa} | j_1 j_2 \rangle \delta_{i_3 j_3} \right] a^\dag_{i_1} a^\dag_{i_2} a^\dag_{i_3} |0\rangle \langle 0 | a_{j_1} a_{j_2} a_{j_3} \\
H_{abb,abb} &=  \sum_{i j \tilde{k} \tilde{p}}\left[ \langle i |T_{a} | j \rangle \delta_{\tilde{k} \tilde{p}}  
+ \langle \tilde{k} |T_{d} | \tilde{p} \rangle \delta_{i j} + E \delta_{i j} \delta_{\tilde{k} \tilde{p}} \right] a^\dag_{i} d^\dag_{\tilde{k}} |0\rangle \langle 0 | a_{j} d_{\tilde{p}} \\
H_{abb,aaa} & = \sum_{i \tilde{k} \underline{j}} \left[ \frac{1}{\sqrt{2}} \langle \chi, \tilde{k} | j_2 j_3 \rangle \delta_{i j_1}  \right]
a^\dag_{i_1} d^\dag_{\tilde{k}} |0\rangle \langle 0 | a_{j_1} a_{j_2} a_{j_3} \\
H_{aaa,abb} & = \sum_{\underline{i} j \tilde{p}} \left[ \frac{1}{\sqrt{2}} \langle i_2 i_3 | \chi, \tilde{k} \rangle \delta_{i_1 j}  \right]
a^\dag_{i_1} a^\dag_{i_2} a^\dag_{i_3} |0\rangle \langle 0 | a_{j_1} d_{\tilde{p}}
\end{align}
or
\begin{align}
H &=  \sum_{i j}\langle i |T_{a} | j \rangle a^\dag_{i}  a_{j} + \sum_{i_1 i_2 j_1 j_2}\frac{1}{2} \langle i_1 i_2 | V_{aa,aa} | j_1 j_2 \rangle  a^\dag_{i_1} a^\dag_{i_2} a_{j_1} a_{j_2} \nonumber \\
& \phantom{=} + \sum_{\tilde{k} \tilde{p}} \langle \tilde{k} |T_{d} + E | \tilde{p} \rangle  d^\dag_{\tilde{k}} d_{\tilde{p}} + \sum_{\tilde{k} j_1 j_2} \left[ \frac{1}{\sqrt{2}} \langle \chi, \tilde{k} | j_1 j_2 \rangle  \right]
 d^\dag_{\tilde{k}}  a_{j_1} a_{j_2} + h.c. \, ,
\end{align}
which is the Hamiltonian usually considered in effective field theory.

\subsubsection{$Q = \underline{ab}$ realization}
As a next example we consider the case 
\begin{equation}
\langle i_{\alpha_1} i_{\alpha_2} |  \frac{\langle \sigma_{\alpha_1}\sigma_{\alpha_2} | - \langle \sigma_{\alpha_2}\sigma_{\alpha_1} | }{\sqrt{2}}  V  \frac{| \sigma_{\beta_1}\sigma_{\beta_2} \rangle - | \sigma_{\beta_2}\sigma_{\beta_1} \rangle }{\sqrt{2}}   | i_{\beta_1} i_{\beta_2} \rangle 
= \langle i_{\alpha_1} i_{\alpha_2} | V_{\sigma_{\alpha_1}\sigma_{\alpha_2}, \sigma_{\beta_1}\sigma_{\beta_2} }^{A_\sigma} | i_{\beta_1} i_{\beta_2} \rangle = 0
\end{equation}
and
\begin{equation}
\langle i_{\alpha_1} i_{\alpha_2} |  \frac{\langle \sigma_{\alpha_1}\sigma_{\alpha_2} | + \langle \sigma_{\alpha_2}\sigma_{\alpha_1} | }{\sqrt{2 + 2 \delta_{\alpha_1 \alpha_2}}}  V  \frac{| \sigma_{\beta_1}\sigma_{\beta_2} \rangle + | \sigma_{\beta_2}\sigma_{\beta_1} \rangle }{\sqrt{2 + 2 \delta_{\beta_1 \beta_2}}}   | i_{\beta_1} i_{\beta_2} \rangle 
= \langle i_{\alpha_1} i_{\alpha_2} | V_{\sigma_{\alpha_1}\sigma_{\alpha_2}, \sigma_{\beta_1}\sigma_{\beta_2} }^{S_\sigma} | i_{\beta_1} i_{\beta_2} \rangle 
\end{equation}
with
\begin{equation}
\langle i_{\alpha_1} i_{\alpha_2} | V_{\sigma_{\alpha_1}\sigma_{\alpha_2}, \sigma_{\beta_1}\sigma_{\beta_2} }^{S_\sigma} | i_{\beta_1} i_{\beta_2} \rangle = 0 \quad \forall \quad \left[(\sigma_{\alpha_1}\sigma_{\alpha_2}) \neq (aa) \text{ or } (ab) \right] \text{or} \left[ (\sigma_{\beta_1}\sigma_{\beta_2}) \neq (aa) \text{ or } (ab) \right]
\end{equation}
and no coupling between the symmetric and antisymmetric spin components.
We then have 
\begin{align}
V_{3b} \mathcal{P}_S^3 
&= \sum_{i_1 i_2 j_1 j_2} \frac{1}{2} \left[
\langle i_1 i_2 | V_{aa,aa} | j_1 j_2 \rangle a^\dag_{i_1} a^\dag_{i_2} a_{j_1} a_{j_2} \right. \nonumber\\ 
& \phantom{=} \left. \qquad \qquad \qquad 
+ \langle i_1 i_2 | V_{aa,ab} | j_1 j_2 \rangle a^\dag_{i_1} a^\dag_{i_2} a_{j_1} b_{j_2}
+ \langle i_1 i_2 | V_{aa,ba} | j_1 j_2 \rangle a^\dag_{i_1} a^\dag_{i_2} b_{j_1} a_{j_2} \right. \nonumber\\ 
& \phantom{=} \left. \qquad \qquad \qquad 
+ \langle i_1 i_2 | V_{ab,aa} | j_1 j_2 \rangle a^\dag_{i_1} b^\dag_{i_2} a_{j_1} a_{j_2}
+ \langle i_1 i_2 | V_{ba,aa} | j_1 j_2 \rangle b^\dag_{i_1} a^\dag_{i_2} a_{j_1} a_{j_2} \right. \\ 
& \phantom{=} \left. \qquad \qquad \qquad 
+ \langle i_1 i_2 | V_{ab,ab} | j_1 j_2 \rangle a^\dag_{i_1} b^\dag_{i_2} a_{j_1} b_{j_2}
+ \langle i_1 i_2 | V_{ab,ba} | j_1 j_2 \rangle a^\dag_{i_1} b^\dag_{i_2} b_{j_1} a_{j_2} \right. \nonumber \\ 
& \phantom{=} \left. \qquad \qquad \qquad 
+ \langle i_1 i_2 | V_{ba,ab} | j_1 j_2 \rangle b^\dag_{i_1} a^\dag_{i_2} a_{j_1} b_{j_2}
+ \langle i_1 i_2 | V_{ba,ba} | j_1 j_2 \rangle b^\dag_{i_1} a^\dag_{i_2} b_{j_1} a_{j_2} \right] \nonumber \\
&= \sum_{i_1 i_2 j_1 j_2} \frac{1}{2} \left[
\langle i_1 i_2 | V_{aa,aa} | j_1 j_2 \rangle a^\dag_{i_1} a^\dag_{i_2} a_{j_1} a_{j_2} \right. \nonumber\\ 
& \phantom{=} \left. \qquad \qquad \qquad 
+ \left(\langle i_1 i_2 | V_{aa,ab} | j_1 j_2 \rangle + \langle i_2 i_1 | V_{aa,ba} | j_2 j_1 \rangle \right) a^\dag_{i_1} a^\dag_{i_2} a_{j_1} b_{j_2} \right. \\ 
& \phantom{=} \left. \qquad \qquad \qquad 
+ \left( \langle i_1 i_2 | V_{ab,aa} | j_1 j_2 \rangle + \langle i_2 i_1 | V_{ba,aa} | j_2 j_1 \rangle \right) a^\dag_{i_1} b^\dag_{i_2} a_{j_1} a_{j_2} \right. \nonumber\\ 
& \phantom{=} \left. \qquad \qquad \qquad 
+ \left( 
\langle i_1 i_2 | V_{ab,ab} | j_1 j_2 \rangle
+ \langle i_1 i_2 | V_{ab,ba} | j_2 j_1 \rangle 
+ \langle i_2 i_1 | V_{ba,ab} | j_1 j_2 \rangle 
+ \langle i_2 i_1 | V_{ba,ba} | j_2 j_1 \rangle 
\right) a^\dag_{i_1} b^\dag_{i_2} a_{j_1} b_{j_2} \right] \nonumber\\
&= \sum_{i_1 i_2 j_1 j_2} \frac{1}{2} \left[
\langle i_1 i_2 | V_{aa,aa} | j_1 j_2 \rangle a^\dag_{i_1} a^\dag_{i_2} a_{j_1} a_{j_2} \right. \nonumber\\ 
& \phantom{=} \left. \qquad \qquad \qquad 
+ 2\langle i_1 i_2 | V_{aa,ab} | j_1 j_2 \rangle a^\dag_{i_1} a^\dag_{i_2} a_{j_1} b_{j_2} \right. \\ 
& \phantom{=} \left. \qquad \qquad \qquad 
+ 2 \langle i_1 i_2 | V_{ab,aa} | j_1 j_2 \rangle a^\dag_{i_1} b^\dag_{i_2} a_{j_1} a_{j_2} \right. \nonumber\\ 
& \phantom{=} \left. \qquad \qquad \qquad 
+ \left( 
 2\langle i_1 i_2 | V_{ab,ab} | j_1 j_2 \rangle
+ 2 \langle i_1 i_2 | V_{ab,ba} | j_2 j_1 \rangle  
\right) a^\dag_{i_1} b^\dag_{i_2} a_{j_1} b_{j_2} \right] \nonumber \\
&= \sum_{i_1 i_2 j_1 j_2} \frac{1}{2} \left[
\langle i_1 i_2 | V_{aa,aa}^{S_\sigma} | j_1 j_2 \rangle a^\dag_{i_1} a^\dag_{i_2} a_{j_1} a_{j_2} \right. \nonumber\\ 
& \phantom{=} \left. \qquad \qquad \qquad 
+ \sqrt{2} \langle i_1 i_2 | V_{aa,ab}^{S_\sigma} | j_1 j_2 \rangle a^\dag_{i_1} a^\dag_{i_2} a_{j_1} b_{j_2} \right. \\ 
& \phantom{=} \left. \qquad \qquad \qquad 
+ \sqrt{2} \langle i_1 i_2 | V_{ab,aa}^{S_\sigma} | j_1 j_2 \rangle a^\dag_{i_1} b^\dag_{i_2} a_{j_1} a_{j_2} \right. \nonumber\\ 
& \phantom{=} \left. \qquad \qquad \qquad 
+ \left( 
 \langle i_1 i_2 | V_{ab,ab}^{S_\sigma} | j_1 j_2 \rangle
+ \langle i_1 i_2 | V_{ab,ab}^{S_\sigma} | j_2 j_1 \rangle  
\right) a^\dag_{i_1} b^\dag_{i_2} a_{j_1} b_{j_2} \right] \nonumber
 \, . 
\end{align}
We can rewrite the Hamiltonian acting in the $aaa$, $aab$ and $abb$ channels as
\begin{equation}
H = H_{aaa,aaa} + H_{aab,aaa} + H_{aaa,aab} + H_{aab,aab} + H_{abb,aab} + H_{aab,abb} + H_{abb,abb} \, ,
\end{equation}
with 
\begin{align}
H_{aaa,aaa} &=  \sum_{\underline{i} \underline{j}}\left[ \frac{1}{2} \langle i_1 |T_{a} | j_1 \rangle \delta_{i_2 j_2} \delta_{i_3 j_3} + \frac{1}{2} \langle i_1 i_2 | V_{aa,aa}^{S_\sigma} | j_1 j_2 \rangle \delta_{i_3 j_3} \right] a^\dag_{i_1} a^\dag_{i_2} a^\dag_{i_3} |0\rangle \langle 0 | a_{j_1} a_{j_2} a_{j_3} \\
H_{aab,aab} &=  \sum_{\underline{i} \underline{j}}\left[ \frac{1}{2} \langle i_3 |T_{b} | j_3 \rangle \delta_{i_1 j_1} \delta_{i_2 j_2} 
+ \langle i_1 |T_{a} | j_1 \rangle \delta_{i_2 j_2} \delta_{i_3 j_3} + \frac{1}{2} \langle i_1 i_2 | V_{aa,aa} | j_1 j_2 \rangle \delta_{i_3 j_3} \right. \nonumber \\
& \phantom{=} \qquad \qquad \left. + \frac{1}{2} \left( 
 \langle i_2 i_3 | V_{ab,ab}^{S_\sigma} | j_2 j_3 \rangle
+ \langle i_2 i_3 | V_{ab,ab}^{S_\sigma} | j_3 j_2 \rangle  
\right) \delta_{i_1 j_1} \right] a^\dag_{i_1} a^\dag_{i_2} b^\dag_{i_3} |0\rangle \langle 0 | a_{j_1} a_{j_2} b_{j_3} \\
H_{abb,abb} &=  \sum_{\underline{i} \underline{j}}\left[ \frac{1}{2} \langle i_1 |T_{a} | j_1 \rangle \delta_{i_2 j_2} \delta_{i_3 j_3} 
+ \langle i_2 |T_{b} | j_2 \rangle \delta_{i_1 j_1} \delta_{i_3 j_3} \right.\nonumber \\
& \phantom{=} \qquad \qquad \left. + \frac{1}{2} \left( 
 \langle i_1 i_2 | V_{ab,ab}^{S_\sigma} | j_1 j_2 \rangle
+ \langle i_1 i_2 | V_{ab,ab}^{S_\sigma} | j_2 j_1 \rangle  
\right) \delta_{i_3 j_3} \right] a^\dag_{i_1} b^\dag_{i_2} b^\dag_{i_3} |0\rangle \langle 0 | a_{j_1} b_{j_2} b_{j_3} \\
H_{aab,aaa} & = \sum_{\underline{i} \underline{j}} \left[ \frac{1}{\sqrt{2}} \langle i_2 i_3 | V_{ab,aa}^{S_\sigma} | j_2 j_3 \rangle \delta_{i_1 j_1}  \right]
a^\dag_{i_1} a^\dag_{i_2} b^\dag_{i_3} |0\rangle \langle 0 | a_{j_1} a_{j_2} a_{j_3} \\
H_{aaa,aab} & = \sum_{\underline{i} \underline{j}} \left[ \frac{1}{\sqrt{2}} \langle i_2 i_3 | V_{aa,ab}^{S_\sigma} | j_2 j_3 \rangle \delta_{i_1 j_1}  \right]
a^\dag_{i_1} a^\dag_{i_2} a^\dag_{i_3} |0\rangle \langle 0 | a_{j_1} a_{j_2} b_{j_3} \\
H_{abb,aab} & = \sum_{\underline{i} \underline{j}} \left[ \frac{1}{\sqrt{2}} \langle i_1 i_2 | V_{ab,aa}^{S_\sigma} | j_1 j_2 \rangle \delta_{i_3 j_3}  \right]
a^\dag_{i_1} b^\dag_{i_2} b^\dag_{i_3} |0\rangle \langle 0 | a_{j_1} a_{j_2} b_{j_3} \\
H_{aab,abb} & = \sum_{\underline{i} \underline{j}} \left[ \frac{1}{\sqrt{2}} \langle i_1 i_2 | V_{aa,ab}^{S_\sigma} | j_1 j_2 \rangle \delta_{i_3 j_3}  \right]
a^\dag_{i_1} a^\dag_{i_2} b^\dag_{i_3} |0\rangle \langle 0 | a_{j_1} b_{j_2} b_{j_3}
\end{align}
\clearpage
This leads us to
\begin{align}
H_{aaa,aaa} &= \left(T_{a}^1 + T_{a}^2 + T_{a}^3 + V_{aa,aa}^{23} + V_{aa,aa}^{31} + V_{aa,aa}^{12} \right) \mathcal{P}_{S,aaa}^3 \\
H_{aab,aab} &=  \left( 1 + P_{12,aab} \right) \sum_{\underline{i} \underline{j}}\left[ \frac{1}{2} \langle i_3 |T_{b} | j_3 \rangle \delta_{i_1 j_1} \delta_{i_2 j_2} 
+ \langle i_1 |T_{a} | j_1 \rangle \delta_{i_2 j_2} \delta_{i_3 j_3} + \frac{1}{2} \langle i_1 i_2 | V_{aa,aa}^{S_\sigma} | j_1 j_2 \rangle \delta_{i_3 j_3} \right. \nonumber\\
\label{eq:aabHamiltonian}& \phantom{=} \qquad \qquad \left. + \frac{1}{2} \left( 
 \langle i_2 i_3 | V_{ab,ab}^{S_\sigma} | j_2 j_3 \rangle
+ \langle i_2 i_3 | V_{ab,ab}^{S_\sigma} | j_3 j_2 \rangle  
\right) \delta_{i_1 j_1} \right] | \underline{i}\rangle_{aab} {}_{aab}\langle \underline{j} | \left( 1 + P_{12,aab} \right)/2 \\
& = \left( 1 + P_{12,aab} \right)\left[ \frac{1}{2} T_{b}^3 
+ T_{a}^1 + \frac{1}{2}  V_{aa,aa}^{S_\sigma, 12} + \frac{1}{2} 
  V_{ab,ab}^{S_\sigma,23} \left( 1 + P_{23,aab} \right) \right] \left( 1 + P_{12,aab} \right)/2
\\
& = \left[ T_{a}^1 + T_{a}^2 + T_{b}^3 
 +  V_{aa,aa}^{S_\sigma, 12} + \left( 1 + P_{12,aab} \right) \frac{1}{2} 
  V_{ab,ab}^{S_\sigma,23} \left( 1 + P_{23,aab} \right) \right] \left( 1 + P_{12,aab} \right)/2
\\
H_{abb,abb} &=  \left( 1 + P_{23,abb} \right) \sum_{\underline{i} \underline{j}}\left[ \frac{1}{2} \langle i_1 |T_{a} | j_1 \rangle \delta_{i_2 j_2} \delta_{i_3 j_3} 
+ \langle i_2 |T_{b} | j_2 \rangle \delta_{i_1 j_1} \delta_{i_3 j_3} \right. \nonumber\\
\label{eq:abbHamiltonian}
& \phantom{=} \qquad \qquad \left. + \frac{1}{2} \left( 
 \langle i_1 i_2 | V_{ab,ab}^{S_\sigma} | j_1 j_2 \rangle
+ \langle i_1 i_2 | V_{ab,ab}^{S_\sigma} | j_2 j_1 \rangle  
\right) \delta_{i_3 j_3} \right]  | \underline{i}\rangle_{abb} {}_{abb}\langle \underline{j} |  \left( 1 + P_{23,abb} \right) /2 \\
& = \left[ T_{a}^1 + 
+ T_{b}^2 + T_{b}^3  + 
 \left( 1 + P_{23,abb} \right) \frac{1}{2} V_{ab,ab}^{S_\sigma,12} \left( 1 + P_{12,abb} \right) \right] \left( 1 + P_{23,abb} \right) /2
\\
H_{aab,aaa} & = \sum_{\underline{i} \underline{j}} \left[ \frac{1}{\sqrt{2}} \langle i_2 i_3 | V_{ab,aa}^{S_\sigma} | j_2 j_3 \rangle \delta_{i_1 j_1}  \right] \sqrt{6} /\sqrt{2} \left( 1 + P_{12,aab} \right) |\underline{i} \rangle_{aab} {}_{aaa}\langle \underline{j} | \mathcal{P}_{S,aaa}^3\\
& =  \frac{\sqrt{3}}{\sqrt{2}} \left( 1 + P_{12,aab} \right) V_{ab,aa}^{S_\sigma,23} \mathcal{P}_{S,aaa}^3
\\
H_{aaa,aab} & = \sum_{\underline{i} \underline{j}} \left[ \frac{1}{\sqrt{2}} \langle i_2 i_3 | V_{aa,ab}^{S_\sigma} | j_2 j_3 \rangle \delta_{i_1 j_1}  \right]
 2 \sqrt{3} \mathcal{P}_{S,aaa}^3 |\underline{i} \rangle_{aaa} {}_{aab}\langle \underline{j} | \left( 1 + P_{12,aab} \right) / 2 \\
& = \frac{1}{\sqrt{6}}\left( 1 + P_+ + P_- \right) \left( 1 + P_{12,aaa}\right) V_{aa,ab}^{S_\sigma,23} \left( 1 + P_{12,aab} \right) / 2
\\
H_{abb,aab} & = \sum_{\underline{i} \underline{j}} \left[ \frac{1}{\sqrt{2}} \langle i_1 i_2 | V_{ab,aa}^{S_\sigma} | j_1 j_2 \rangle \delta_{i_3 j_3}  \right]
\left( 1 + P_{23,abb} \right) |\underline{i} \rangle_{abb} {}_{aab}\langle \underline{j} | \left( 1 + P_{12,aab} \right)/2 \\
& = \left( 1 + P_{23,abb} \right) \frac{1}{\sqrt{2}}  V_{ab,aa}^{S_\sigma,12} \left( 1 + P_{12,aab} \right)/2
\\
H_{aab,abb} & = \sum_{\underline{i} \underline{j}} \left[ \frac{1}{\sqrt{2}} \langle i_1 i_2 | V_{aa,ab}^{S_\sigma} | j_1 j_2 \rangle \delta_{i_3 j_3}  \right]
\left( 1 + P_{12, aab} \right) |\underline{i} \rangle_{abb}{}_{abb}\langle \underline{j} | \left( 1 + P_{23, abb} \right)/2 \\
& = \left( 1 + P_{12, aab} \right) \frac{1}{\sqrt{2}}  V_{aa,ab}^{S_\sigma,12} \left( 1 + P_{23, abb} \right)/2 \, .
\end{align}
Note that even when the couplings between the channels are of separable form, it is not straightforward to rewrite this Hamiltonian in terms of a dimer field.
However, since we can interpret our results in terms of couplings to a closed-channel trimer state maybe the introduction of a trimer field could lead to a good approximation.

\end{widetext}

\bibliographystyle{apsrev4-1}
\bibliography{biblio-TS-DJM,biblio-TS}

\end{document}